\documentclass[conference]{IEEEtran}
\usepackage{changes}
\usepackage{epstopdf}
\usepackage
{hyperref}  
\usepackage{enumerate}
\usepackage
{color} 
\usepackage{booktabs} 
\usepackage[T1]{fontenc}
\usepackage{enumitem}
 \usepackage{dutchcal}
\usepackage{fancybox}
\usepackage{graphicx}
\usepackage{wrapfig}
\usepackage[english]{babel}
\usepackage[utf8x]{inputenc}
\usepackage{lipsum}
\usepackage{lineno} 
\usepackage{hyperref}
\usepackage[misc,geometry]{ifsym}
\usepackage{graphicx}
\usepackage[normalem]{ulem}
\usepackage{ucs}
\usepackage{url}
\usepackage{amsbsy}
\usepackage[nointegrals]{wasysym}
\usepackage{amsfonts}
\usepackage{amsmath}
\usepackage{amstext}
\usepackage{amssymb}

\usepackage{xspace}
\usepackage{wrapfig}
\usepackage{subfigure}
\usepackage{algorithm}
\usepackage{verbatim}
\usepackage{multirow}

\usepackage{rotating}
\usepackage{lscape}
\usepackage{epsfig}
\usepackage{float}
\usepackage{mathrsfs}
\usepackage{array}
\usepackage{latexsym}
\usepackage{xspace}

\usepackage{cite}

\usepackage{amsmath,amssymb,amsfonts}
\usepackage{textcomp}
\usepackage{xcolor}

\newlength{\cellH}
\newlength{\cellW}

\definechangesauthor[name={Zinovy}, color=blue]{zd}
\definechangesauthor[name={Nick}, color=green]{na}

\newcommand{\mynote}[2]{
    \fbox{\bfseries\sffamily\scriptsize#1}
    {\small$\blacktriangleright$
    \textcolor{blue}{\textsf{\emph{#2}}}
    $\blacktriangleleft$}}

\newcommand\zd[1]{\mynote{ZD}{#1}}

    \newcommand\mymarginNew[2]{%
    	\marginpar{\vspace{#2}\flushleft\textit{\footnotesize\textcolor{blue} {#1}}}}

\newcommand\hlightNew[3]{%
	\textcolor{blue} {#1} 
	{\mymarginNew{#2}{#3} 
}}

\newcommand\nahll[3]{\hlightNew{#1}{NA:~#2}{#3}}

\newcommand\zdnewwok[3]{{#1}}
\newcommand\zdnewwno[3]{}

\newcommand\nanewok[2]{{#1}}
\newcommand\nanewno[2]{}

\newcommand\awmodok[3]{\modifyok{#1}{#2}{AW:#3}}
\newcommand\awmodno[3]{\modifyno{#1}{#2}{AW:#3}}

\newcommand\mlmodok[3]{\modifyok{#1}{#2}{ML:#3}}
\newcommand\mlmodno[3]{\modifyno{#1}{#2}{ML:#3}}

\newcommand\zdmodok[3]{\modifyok{#1}{#2}{ZD:#3}}
\newcommand\zdmoddok[3]{\modiffyok{#1}{#2}{ZD:#3}}

\newcommand\zdmoddno[3]{\modiffyno{#1}{#2}{ZD:#3}}

\newcommand\namod[3]{\modify{#1}{#2}{NA:~#3}}

\newcommand\namodok[3]{\modifyok{#1}{#2}{NA:#3}}
\newcommand\namoddok[3]{\modiffyok{#1}{#2}{NA:#3}}
\newcommand\namodno[3]{\modifyno{#1}{#2}{NA:#3}}

\newcommand{\modifyok}[3]{{#2}}
\newcommand{\modiffyok}[4]{{#2}}

\newcommand{\modifyno}[3]{{#1}}
\newcommand{\modiffyno}[4]{{#1}}

\newcommand\newno[2]{}

\newcommand\df{dataflow}
\newcommand\asson{association}

\newcommand\refin{refinement}
\newcommand\req{requirement}

\newcommand\multy{multiplicity}
\newcommand\multi{multipliciti}

\newcommand\eg{e.g.}
\newcommand\ie{i.e.}

\newcommand\etc{etc.}

\newcommand\apriori{{\em a priori}}

\newcommand\trafon{transformation}

\newcommand\corring{corresponding}

\newcommand\fwk{framework}

\newcommand\comp{\ensuremath{\,{\triangleright\!\triangleright}\,}}

\newcommand\figrii{}

\newcommand\respace{\!}
\newcommand\lsemm{\left[ \respace\left[ }
\newcommand\rsemm{\right]\respace\right]}
\newcommand\semm[1]{\mbox{$\lsemm\, \mathit{#1} \, \rsemm$}}
\renewcommand\semm[1]{\ensuremath{\lsemm{#1} \rsemm}}

\DeclareMathAlphabet{\mathantt}{OT1}{antt}{li}{it}
\DeclareMathAlphabet{\mathpzc}{OT1}{pzc}{m}{it}

\newtheorem{remark}{Remark}

\newcommand\figref[1]{Fig.~\ref{#1}}

\newcommand\defref[1]{Def.~\ref{def:#1}}

\renewcommand\eqref[1]{\ref{eq:#1}}
\newcommand\sectref[1]{Sect.~\ref{sec:#1}}
\newcommand\secref[1]{Sect.~\ref{sec:#1}}

\makeatletter
  \def\fps@figure{tp}
\makeatother

\newcounter{defCounter}
\newenvironment{defin}[1][]{
\refstepcounter{defCounter}
\begin{trivlist}
\item[\hskip \labelsep {\bfseries Definition \arabic{defCounter} (#1) }]}%
{\end{trivlist}}

\newcommand\asson{association}

\newcommand\refin{refinement}

\newcommand\eg{e.g.}
\newcommand\ie{i.e.}

\newcommand\etc{{\em etc}}

\newcommand\apriori{{\em a priori}}

\newcommand\respace{\!}
\newcommand\lsemm{\left[ \respace\left[ }
\newcommand\rsemm{\right]\respace\right]}

\newcommand\dolan
{\mbox{$\Delta $}}

\newcommand\mathsymbol[1]{\mbox{$#1$}}
\newcommand\world
{\mathsymbol{\cal W}}
\newcommand\cmter
{\mathsymbol{\cal C}}

\newcommand\df{\formalPar{D}}

\newcommand\formalObj[1]{\mbox{\sffamily #1}}

\newcommand\informalObj[1]{\mbox{\textit #1}}

\newcommand\FormalInterpret[1]{\mbox{$\formalObj{i}_f$}}
\newcommand\SubstantInterpret[1]{\mbox{$\informalObj{i}_s$}}

\newcommand\formalPar[1]{\mbox{$\mathrm{#1}$}}

\renewcommand\eg{e.g.}

\newcommand\der[2]{\indexedoperator{der}{#1}{#2}}

\newcommand\ver{\vect{R}}

\newcommand\val{\keyword{value}}

\newcommand\newrevem[1]{\mbox{$\semm{\mmmapname{#1}}$}}

\newcommand\req{\newrevem{q}}

\newcommand\incrar[3] {\mbox{$#1\!:\:#2\:\hookrightarrow\:#3$}}

\newcommand\flar[3]{\ensuremath{#1\!:#2\leftarrow #3}} 

\usepackage{diagrams}

\newarrowhead{bt}\blacktriangleright\blacktriangleright\blacktriangleright\blacktriangleright
\newarrowtail{bt}\blacktriangleleft\blacktriangleleft\blacktriangleleft\blacktriangleleft
\newarrowmiddle{bar}\rtbar\ltbar\dtbar\utbar
\newarrowmiddle{|}\vert\vert--
\newarrowmiddle{||}\Vert\Vert==
\newarrowmiddle{=}==\Vert\Vert
\newarrowtail{=}==\Vert\Vert
\newarrowhead{l}\langle\langle\langle\langle 
\newarrowhead{r}\rangle\rangle\rangle\rangle 
\newarrowfiller{=}==\Vert\Vert
\newarrowfiller{d}\cdot\cdot\cdot\cdot
\newarrowmiddle{x}****
\newarrowmiddle{b}\bullet\bullet\bullet\bullet
\newarrowtail{b}\bullet\bullet\bullet\bullet
\newarrowmiddle{3}\equiv\equiv\vfthree\vfthree
\newarrowfiller{3}\equiv\equiv\vfthree\vfthree
\newarrowtail{3}\equiv\equiv\vfthree\vfthree
\newarrowtail{<=}\Leftarrow\Rightarrow{\@cmex7E}{\@cmex7F}
\newarrowfiller{bold}{\bf-}{\bf-}{\bf|}{\bf|}  
\newarrowfiller{o}{\hho}{\circ}{\circ}{\circ}  
\newarrowmiddle{>}\rtla\ltla\dtla\utla
\newarrowmiddle{d}\cdot\cdot\cdot\cdot

\newarrow{To}----{>}
\newarrow{Mapsto}b---{>}
\newarrow{Maps}b----
\newarrow{ID}33333

\newarrow{EQ}=====
\newarrow{NRelto}--+-{->}
\newarrow{Relto}--b-{->}
\newarrow{BSpanto}--b-{->} 
\newarrow{Embed}>---{->}
\newarrow{Dashto}{}{dash}{}{dash}{>} 
\newarrow{RDiagto}3333{r}
\newarrow{RDiagderto}{}{3}{}{3}{r}
\newarrow{LDiagto}3333{l}
\newarrow{LDiagderto}{}{3}{}{3}{l}
\newarrow{Mapsderto}{b}{dash}{}{dash}{>}
\newarrow{Into}C---{>}
\newarrow{Indashto}C{dash}{}{dash}{>}
\newarrow{Derinto}C{dash}{}{dash}{>}
\newarrow{Congruent}33333
\newarrow{Cover}----{blacktriangle}
\newarrow{Monic}{>}--->
\newarrow{Isoto}{>}---{triangle}
\newarrow{ISA}===={=>}
\newarrow{Isato}===={=>}
\newarrow{Doubleto}===={=>}
\newarrow{ClassicMapsto}{|}--->
\newarrow{Entail}{|}----
\newarrow{PArrow}{o}--->
\newarrow{MArrow}----{>>}
\newarrow{MPArrow}{o}---{>>}
\newarrow{Ogogoto}oooo{->}
\newarrow{Metato}--3->
\newarrow{MetaMapto}{|}-3->
\newarrow{OneToMany}+---{o}
\newarrow{HalfDashTo}{}{dash}{}-{>>}
\newarrow{DLine}=====
\newarrow{Line}-----
\newarrow{Tline}33333
\newarrow{Dashline}{}{dash}{}{dash}{}
\newarrow{Dotline}{}{o}{}{o}{}
\newarrow{Curlyto}{curlyvee}--->           
\newarrow{Bito}<--->
\newarrow{Bito}{<}---{>}
\newarrow{Bidito}{<}==={>}
\newarrow{Bitrito}{bt}333{bt}
\newarrow{Corrto}<--->
\newarrow{Dercorrto}{<}{dash}{}{dash}{>}

\newarrow{Instofto}{curlyvee}...{>}
\newarrow{Instofderto}{curlyvee}{dash}{}{dash}{->}
\newarrow{Updto}----{>}
\newarrow{Derupdto}{}{dash}{}{dash}{>}
\newarrow{Mchto}----{>}
\newarrow{Dermchto}{}{dash}{}{dash}{>}
\newarrow{Viewto}----{>->}
\newarrow{Derviewto}{}{dash}{}{dash}{>>}
\newarrow{Viewto}===={=>}
\newarrow{Hetmchto}===={=>}
\newarrow{Derviewto}{=}{}{=}{}{=>}
\newarrow{Idleto}===={=>}
\newarrow{Deridleto}{=}{}{=}{}{=>}

\newlength{\StatArrBody}
\newlength{\NodeFrameThickness}
\newarrow{Updto}----{>}
\newarrow{Derupdto}{}{dash}{}{dash}{>}
\newarrow{Mchto}----{>}
\newarrow{Dermchto}{}{dash}{}{dash}{>}
\newarrow{Uto}----{>}
\newarrow{Uderto}{}{dash}{}{dash}{>}
\newarrow{Mto}----{>}
\newarrow{Mderto}{}{dash}{}{dash}{>}

\newcommand\diagopername[1]{\ensuremath{\mathsf{#1}}}

\newcommand\putl[1]{\mbox{\diagopername{put}$_{#1}$}} 
\newcommand\putlback[1]{\ensuremath{\overleftarrow{\putl[1]}}}
\renewcommand\putl{\mbox{\diagopername{put}}} 
\renewcommand\putlback{\ensuremath{\overleftarrow{\putl}}}
\renewcommand\putlback{\ensuremath{\putl^\circlearrowleft}}

\newcommand\given{\nmf{given}}
\newcommand\der{\nmf{der}}

\newcommand\val{\nmf{valid}}
\newcommand\Mit{\nmf{Mit}}
\newcommand\Elim{\nmf{Elim}}
\newcommand\True{\nmf{True}}

\newcommand\valid{\nmf{valid}}

\newcommand\ver{\nmf{ver}}

\newcommand\gpr{\nmf{gPr}}
\newcommand\hpr{\nmf{hPr}}

\newcommand\ctt{\ensuremath{\nmf{Cnt}}}

\newcommand\Contt{\nmf{Containment}}
\newcommand\gProtn{\nmf{gr}\Protn}
\newcommand\hProtn{\nmf{hz}\Protn}
\newcommand\hPrtn{\nmf{hz}\Prtn}
\newcommand\Protn{\nmf{Protection}}

\newcommand\grProtn{\nmf{gr}\Protn}
\newcommand\Prtn{\nmf{Protection}}

\newcommand\namefont[1]{\ensuremath{\mathsf{#1}}}
\newcommand\VV{\nmf{V\&V}}
\newcommand\nmf[1]{\namefont{#1}}
 
\newcommand\indfont[1]{\ensuremath{\mathsf{#1}}}

\newcommand\adv{\nmf{adv}}

\newcommand\main{\nmf{main}}
\newcommand\entry{\nmf{entry}}

\newcommand\exe{\indfont{exe}}

\renewcommand\implies{\Rightarrow}

\newcommand\exgsn{Ex.\gsn}
\newcommand\exwfp{Ex.\wfp}
\newcommand\exg{\exgsn}
\newcommand\exw{\exwfp}
\newcommand\exgoal{\ensuremath{Z}}

\newcommand\gsn{\ensuremath{\mathsf{GSN}}}
\newcommand\GSN{\ensuremath{\mathsf{GSN}}}

\newcommand\wpd{\wfp-diagram}
\newcommand\gsd{GSN-diagram}
\newcommand\wtog{\ensuremath{\mathsf{wf}\mathsf{2gsn}}}

\newcommand\enp{\ensuremath{\mathsf{ep}}}

\newcommand\implBy{\ensuremath{\Leftarrow}}

\newcommand\wfinst{\ensuremath{\wfsep^{\exe}}}
\newcommand\wfsepdef{\ensuremath{\wfsep}}

\newcommand\confcheck[2]{\ensuremath{\Vdash_{{#1}}^{{#2}} }}
\newcommand\instan{\nmf{inst}}
\newcommand\gbf{\ensuremath{{\mathbf{G}}}}
\newcommand\csetx[2]{\ensuremath{{\mathbf{C}^{#1}_{#2}} }}
\newcommand\reff{\nmf{ref}}
\newcommand\wfbf{{\bf\em WF$\pmb{^+}$}}

\newcommand\WFP{\ensuremath{\mathsf{WF^+}}}

\newcommand\wfp{\WFP}
\newcommand\wfc{\wfp}

\newcommand\wcfa{\wfp}

\newcommand\numitem[1]{{\bf (#1)}}
\newcommand\numitempar[1]{{\smallskip\noindent \bf (#1)}}

\newcommand\dbox[1]{\fbox{$#1$}}
\renewcommand\dbox[1]{\ovalbox{$#1$}}

\renewcommand\implies{\ensuremath{\Rightarrow}}

 \newcommand\secdercon{\sectref{isnt-metamod}}
\newcommand\exeinst{\ensuremath{\wfexe(X)}}
\newcommand\wfind[1]{\ensuremath{\mathit{Wf}_{#1}}}
\newcommand\wfindd[2]{\ensuremath{\mathit{Wf}_{#1}^{#2}}}
\newcommand\sep{\nmf{SEP}}
\newcommand\norm{\nmf{Norm}}
\newcommand\wfsep{\wfind{\sep}}
\newcommand\wfsepexe{\wfindd{\sep}{\exe}}
\newcommand\wfexe{\wfsepexe}
\newcommand\wfexex{\wfindd{\sep}{\exe}(X)}
\newcommand\wfnorm{\wfind{\norm}}
\newcommand\wfnormmeta{\wfind{\nmf{Metanorm}}}

\newtheorem{defin}{\bf\em Definition}

\newcommand\citegsn{\cite{gsn11}}
\newcommand\citegsnp[1]{\cite[#1]{gsn11}}
\newcommand\citegsN[1]{\cite{gsn18}}
\newcommand\citegsNp[1]{\cite[#1]{gsn18}}

\newcommand\citealan{\cite{alan-caseAgainst-10}}
\newcommand\citedebate{\cite{val-shonan}}

\newcommand\papertr[2]{#1}  

\papertr{}
{
	\usepackage{MCSTR}
	\MCSTRtitle{Assurance via Workflow$^+$  Modelling and Conformance}
	\MCSTRauthor{Zinovy Diskin,  Nicholas Annable, Alan Wassyng
		\\ Reviewed by Mark Lawford, Tom Maibaum}
	\MCSTRauthor{---}
	\MCSTRdate{April 2019}
	\MCSTRid{TR 2019-04-29}
}

\newcommand\paperGM[2]{#1}  

\begin{document}

\papertr{}
{
	\MCSTRmakecover %
}

\title{Assurance via Workflow$^+$  Modelling and Conformance
 }\thanks{Paper deadline is April 29
 “Foundations” and “Practice and Innovation” papers must not exceed 10 pages (including figures and appendices, excluding references). All papers may include 1 additional page for references.}

\author{%
	\IEEEauthorblockN{Zinovy Diskin, Nicholas Annable, Alan Wassyng, Mark Lawford}
\IEEEauthorblockA{\textit{Computing and Software}, McMaster University, Canada}
\\ [-10pt]
\{diskinz, annablnm, wassyng, lawford\}@mcmaster.ca
}

\maketitle
%
\begin{abstract}
We propose considering assurance as a model management enterprise: saying that a system is safe amounts to specifying three workflows modelling how the safety engineering process is defined and executed, and checking their conformance. These workflows are based on precise data modelling as in functional block diagrams,  but their distinctive feature is the presence of relationships between the output data of a process and its input data; hence, the name ``Workflow Plus''.

A typical \WFP-model comprises three layers: (i) process and control flow, (ii) dataflow (with input-output relationships), and (iii) argument flow or constraint derivation. Precise dataflow modelling signifies a crucial distinction of \WFP-based and GSN-based assurance, in which the data layer is mainly implicit. We provide a detailed comparative analysis of the two formalisms and conclude that GSN does not fulfil its promises. 
%
\end{abstract}
%



\section{Introduction} 

\awmodok{Nowadays, safety}{Safety}{} cases (SC) are an established approach to certifying safety-critical systems in \awmodok{automotive, avionics, railway, nuclear and other}{many}{} domains\cite{adelard-mss10}. However, the size and complexity, and hence the cost, of a typical SC for a modern \nanewok{computer-controlled}{} system can be enormous and, in some cases, greater than the cost of building the system \awmodok{as such}{}{} \cite{multileg-bloomf-dsn03}. This explains the long-standing demand for tooling to assist and automate the building and maintenance of \awmodok{SC}{SCs}{}
\cite{denney2018,SCT,ASCE}. The activity in this domain is mainly based on GSN \citegsn\ or CAE \cite{cae15}\ diagrams (further collectively referred to as GSN-style diagrams), %
which \awmodok{became}{has become}{} a de facto standard for \awmodok{structuring and specifying}{documenting}{} SCs.
An OMG initiative in this direction, the Structured Assurance Case Metamodel (SACM)\cite{SACM2}, can also be seen as an effort to standardize \awmodok{the GSN}{GSN-like}{} notation \awmodno{in MOF/UML terms.}{.}{} 



{
In this paper, we report on several shortcomings of the GSN formalism for building and maintaining SCs. \mlmodok{They follow from the following two major drawbacks.}{We will show that the shortcomings are the result of two main GSN's limititations:}{}  a) GSN conducts logical inference that (we will show) essentially involves graph-based object-oriented (GBOO) data, in a significantly  data-simplified (and at some points, data-ignorant) way. b) GSN manages the basic evidence layer of the argument flow, which (we show) is also a GBOO data enterprise, in an unstructured discrete setting leaving an array of important relationships implicit. We demonstrate with an example (taken from the GSN Community Standard \cite{gsn18}) that these drawbacks can essentially distort the argument flow and thus create a dangerous false sense of confidence. A significant investment of time and money into the development of tools for GSN-style SCs may be a misguided endeavour as GSN-based tools will}
\zdmodok{never escape} 
{suffer from}{} the drawbacks of GSN.
We see the need for an assurance \fwk\ alternative to GSN, which will better achieve GSN's 
\zdmodok{original goal of basing assurance on structured reasoning and evidence.}{%
original focus on structure, decomposition, and argument
}. 




We thus return to first principles by specifying a set of requirements \awmodok{to}{for}{} a foundational  \fwk, on which tooling should be based. There are two obvious major requirements necessary to reduce the cost of \awmodok{SCs building and maintenance:}{building and maintaining SCs.}{} 
 \begin{enumerate}[label={\bf  R\arabic*)}, 
	wide=0pt]
\item Provide support for \awmodno{a careful reusability analysis with multiple reusability layers identified.}
{a reusability analysis that encourages an \emph{information hiding approach}~\cite{} as well as some ability to cope with unforeseen changes.}{} 
\item Allow for an incremental approach based on extensive and effective traceability \awmodok{mechanism inside layers (horizontally) and spanning the layers (vertically).}{mechanisms.}{} 
\end{enumerate}
~~These tasks are typical for MDE,  and known to be difficult, but in the assurance context they become especially challenging due to the following \awmodok{factors or {\em Challenges} concisely formulated in}{\emph{challenges}}{} \cite{nasaTR}:    
 \begin{enumerate}[label={\bf  C\arabic*)}, 
 	wide=0pt]
 \item the heterogeneity of the artifacts to be integrated in the safety argument flow
 \item with {the need to include} informal and domain-specific artifacts, 
and a multitude of implicit assumptions about system \awmodok{functioning}{behaviour}{} and interaction;  
\item the necessity to coordinate the functional system development and its safety \awmodok{engineering side.}{engineering.}{} 
\end{enumerate}
{\renewcommand\numitempar[1]{}
 ~~Addressing these challenges gives rise to the following additional requirements. 
   \begin{enumerate}[start=3,label={\bf R\arabic*)}, 
  	wide=0pt]
  	  \item Provide a unified approach to heterogeneous artifacts and argument flow. 
  	\item Be easily understandable by the stakeholders to allow for \awmodok{an}{}{} effective communication, and 
  	be easily extensible to accommodate \awmodok{its results.}{resulting changes from such communication.}{} 
  	\item Be system-engineering friendly.
  	\item \awmodok{Be friendly to safety engineering traditions.}{Be safety-engineering friendly, including the necessity to observe reasonable conservatism}{}
  	 \end{enumerate}
}  
Requirements R3-5 address \awmodok{challenge}{challenges}{} C1-3 resp. (and R5 actually spans across all the challenge range); we added R6 as especially important in the domain of safety-critical systems, where a reasonable conservatism is a forced necessity \citealan. 

In this paper, we propose a modelling \fwk\ referred to as {\em Workflow$^+$} ({briefly, \wfp})which meets the \req s on the specification level and shows promising potential in \awmodok{our}{}{} ongoing collaboration with \awmodok{automotive industry; implementation is}{our automotive partners. Tool implementation has}{} not started yet. The goal of the paper \awmodok{seen broadly}{}{} is to lay theoretical foundations for applying established MDE methods \awmodok{and tools for assurance, and provide the latter with rigour, conceptual and technological clarity, and tool support.}{to the assurance of safety-critical software intensive systems. We believe that this can lead to improved rigour, conceptual and technological clarity, and advanced tool support for both development and assurance of such systems.}{}

\zdmoddok{
\wfp\ employs the idea of a process as a unifying foundation: a car is a process that consumes fuel and produces velocity, and a SEP (safety engineering process) is a process that takes that car's data as its input and produces a set of safety measures and a set of work products demonstrating that ``the car is acceptably safe'' if those measures are implemented. Each of these \awmodok{root}{}{} processes is decomposed into smaller processes (engine, powertrain, wheels etc for the car, and hazard analysis (HA), requirements, design, verification, etc. for SEP) until their complexity becomes manageable. Thus, the actual core of the 
\fwk\ is the notion of workflow: a network of processes working together to satisfy the top-level requirements. 
}{
\wfp\ employs the idea of a process as a unifying foundation for modelling and analysis of both  cyber-physical systems and SEPs (safety engineering processes) that include verification and validation (\VV) of the former analysis.  
}{}{-7em}
We refer to the \fwk\ as \wfp\ due to a special feature of a typical SEP subprocess: its output data are related to its input data rather than disjoint \mlmodok{to}{from}{} them as in the classical block diagram setting. The name \wfp\ refers to both a workflow composed from processes with input-output related data, and the approach to SC development based on such workflows. 

In the following, we describe the content of the paper and list its technical contributions. 
In \sectref{assr-as-conf}, we give 
a formalized definition of assurance based on the notions of \wfp-conformance and refinement. We are not aware of such model-based assurance definitions in the \awmodok{literature; perhaps, their absence had been a major obstacle in providing MDE foundations for assurance.}{literature.}{} 
In \sectref{gsn2wf1}-\ref{sec:gsn2wf2}, we show how the \wfp-method works with an \mlmodok{elaborated}{}{} example: we take a GSN diagram from the GSN Community 
Standard~\citegsN{},  \awmodok{interpreted}{interpret it}{} as an SC, recover its semantics, rearrange it in \wfp-terms, and rebuild the case as \wfp-conformance. We show 
that the argument flow---a focal point of the GSN-based assurance---is a data-driven enterprise  (technically, the inference of derived OCL constraints over data they span from given OCL).
%
This significantly differentiates \wfp\ from GSN, in which data and dataflow are mainly implicit. 
\sectref{anti-gsn} is a comparative analysis of \wfp\ vs. GSN: it can be seen as an evaluation of the new \fwk\ against an established one based on the standard example. We show that the argument flow in the example \awmodok{is significantly distorted and actually wrong}{is significantly distorted---a disappointing}{} observation for an example residing in a popular standard since 2011, Version 1  \citegsnp{p.5} until now in Version 2 \citegsNp{p.14}.
 In \sectref{compara-survey}, we discuss how well \wfp\ addresses the issues described in the empirical survey \cite{jcheng-corr18} and satisfy our \req s above. Section \ref{sec:relwork} \mlmodok{briefly observes}{is a brief overview of}{} related work. \sectref{concl} concludes. 

We extensively use colours in our diagrams as a visually convenient additional way to convey semantics of diagram elements, which is encoded by labels (specifically, UML's stereotypes) and shapes of elements. 
\papertr{
\wfp-diagrams contain large amounts of information and normally would span a full page but we made them smaller than  desired due to space limitations. The accompanying technical report  includes bigger versions of these diagrams, and two additional fragment of the material discussed in \sectref{compara-survey} but omitted due to space limits.
}
{}


\renewcommand\figrii{\figref{fig:recoverSem}}
\section{Assurance as conformance}\label{sec:assr-as-conf}

In this section we introduce our general model of assurance. 

\subsection{Background}
Approaches to certification of safety-critical software systems \awmodok{can roughly be divided in two}{have traditionally been divided into two (overlapping)}{} groups: process-based and product-focused (cf. \cite{alan-caseAgainst-10}). An obvious plus of the former is the clarity and \awmodok{doability}{practicality}{} of the verification: the regulator can always check whether everything prescribed was really executed. An obvious minus is indirectness: if even everything prescribed was executed, \awmodok{it's}{it is}{} unclear how much it guarantees that the product as such is safe. The situation is reversed for \awmodok{the}{}{}product-focused assurance, which assumes that safety can be specified as a collection of attributes such that having their measurable values within a prescribed range can be interpreted as having sufficient evidence that the system is safe. If such a collection could be found, it would provide quite a direct approach to assurance but, unfortunately, reducing safety to an \apriori\ defined finite collection of measurable attributes is unrealistic for complex systems (unless magic creatures like Laplace's daemon are invoked). Indeed, the qualitative and quantitative infinity of possible interactions of such a system as a modern car with its environment (other cars, pedestrians, weather and road conditions, and the driver) makes any hazard analysis performed with a finite number of models and simulations within a finite time period potentially incomplete.

Moreover, any attempt to mitigate this problem with super-intelligent 
mathematical models amenable to super-powerful analysis techniques is doomed to fail as even a perfect ``proof'' \awmodok{(read ``a body of evidence'') of}{of a}{} system's safety is necessarily carried out within some reasoning system $R$ (be it a logic, a math model, a set of assumptions, or a conceptual \fwk), while the adequacy of $R$ to the reality being modeled cannot be established within $R$ and needs an external validation $R'$, for which the story is repeated again.%
\footnote{For example, the celebrated G\"odel's incompleteness theorem shows that even such a poor and seemingly formalizable  universe as arithmetic cannot be formalized in a complete way.} Such properties as safety are inherently probabilistic: only if a sufficiently big number of instances of a concrete product  $X$ is exploited for a sufficiently long period of time in \mlmodok{a}{}{} sufficiently diverse environmental situations, can we then conclude that $X$ is acceptably safe with enough confidence. Unfortunately, when a new system goes into production, neither of the above ``if''s is available so that risk and uncertainty are inherited qualities of a new system. To manage them, classical engineering has always \awmodok{been exercising}{exercised}{} care, incremental innovation and a reasonable conservatism (cf. \citealan). \zdnewwok{However, the latter seems to be \awmodok{not an}{a difficult}{} option at \mlmodno{the}{this}{} time \mlmodno{of}{for}{} autonomous driving, smart homes, the Internet of things and similar initiatives; 
we need a new approach reconciling conservatism and novelty.}{slightly edited}{-3em}

\subsection{Main assumptions.}

Our approach to assurance can be seen as an integration of process- and product-oriented approaches. To make it doable and conservative, we \awmodok{stand}{rely}{} on \awmodok{the}{}{} process-oriented foundations, but \awmodok{we make the}{our}{} notion of process 
includes important aspects of product assurance by including relevant checks on product items using predefined acceptance criteria \cite{iso26262}.
\begin{defin}[Main]\label{def:main}
	\newcommand\gap{\mbox{~~~}}
	System $X$ is considered to be {\em acceptably safe\footnote{The safety aspect is discussed in more detail in Sect.~\ref{sec:safe-systems}.} (for a given application $A$ {in} a given environment $E$)} if the manufacturer's safety engineering process, SEP, satisfies the following two conditions:
\zdnewwok{
	\\
	\gap a)  SEP is acceptably-well defined for a class of systems similar to $X$;  
	\\
	\gap b) SEP's definition is acceptably-well executed for $X$.
	\\	
In other words, according to empirical evidence and expert opinion encoded in condition a), if condition b) is satisfied for a system $X$, then $X$ is safe for application $A$ in environment $E$. 
	}{new phrasing}{-6em}
\end{defin}

\smallskip
To make these general conditions technical and verifiable, we build three \wfp-models: \wfsep\ that models the SEP, $\wfexe(X)$ that specifies SEP's execution for system $X$, and \wfnorm\ that models a body of normative documents (standards, regulations, on-site manuals, expert opinion and all relevant domain knowledge), which regulate and prescribe how SEP is to be defined (but should give the manufacturer enough freedom to adapt to the norm). We specify these models as specially profiled class diagrams (we say metamodels) so that condition a) of \defref{main} amounts to checking if metamodel \wfsep\ is a correct \refin\ of another (normative) metamodel \wfnorm, and condition b) amounts to checking conformance of an object model (instance) $\wfsepexe(X)$ to the metamodel \wfsep. 
%
\begin{defin}[Main, more formal]\label{def:mainF}
	System $X$ is considered to be {\em acceptably safe} if the following two conformance conditions are satisfied: 
\begin{equation}\label{eq:mainF} 
\wfexe(X)\confcheck{\instan}{\VV}\,\wfsep\, \confcheck{\reff}{\VV} \,\wfnorm
\end{equation}
where \confcheck{\instan}{} denotes instanceOf-relation and \confcheck{\reff}{} is a suitably defined refinement relation between workflows modelled as class diagrams. The index \VV\ refers to the necessity to check both Validity of the data  and Verification of the formal relation $\Vdash$; later we will explain how validity checks can be reduced to another layer of syntactic verification.  
\end{defin}
\begin{figure}
\centering
    \includegraphics[
                    width=.475\textwidth%
                 ]%
                 {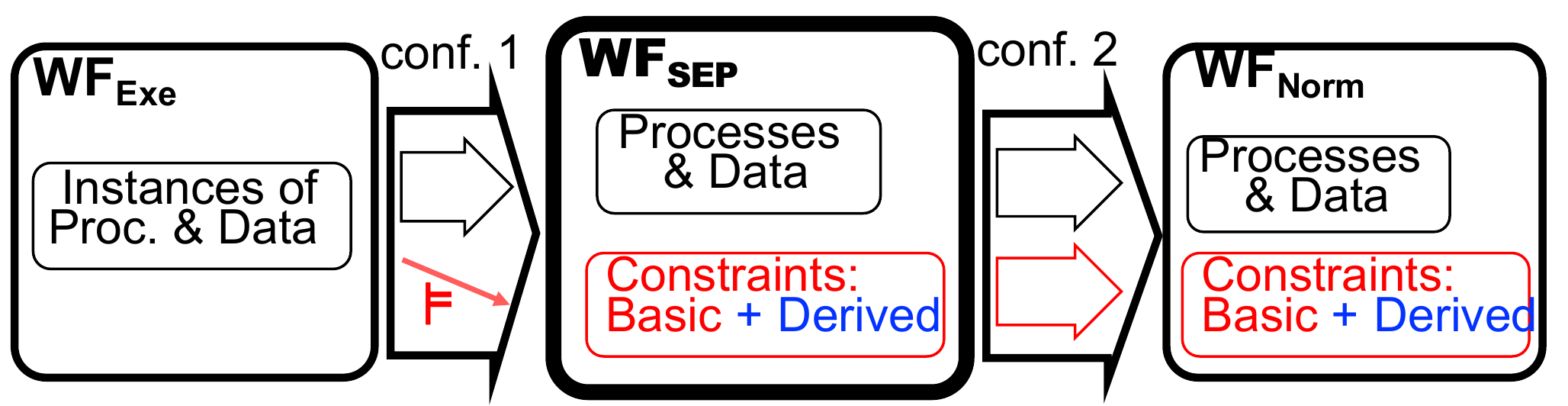}
\caption{Conformance chain} \label{fig:conf-chain}
\end{figure}
Figure~\ref{fig:conf-chain} provides some details. Workflow metamodels \wfsep\ and \wfnorm\ are class diagrams consisting of process classes (\ie, classes modelling processes), data classes, and dataflow associations, and constraints (basic and derived) that force the required semantics of the class diagram. Refinement amounts to having a syntactically correct mapping of the graph of process and data classes and associations in \wfsep\ to a similar graph of \wfnorm, moreover, we require the mapping to be compatible with the constraints. For making the definition more precise and formal, finding an appropriate notion of workflow refinement and its formalization are crucial; we leave  this for future work. As a possible starting point, we are investigating the notion of model transformation \refin\ proposed in \cite{me-models18}; a formal definition of mapping's compatibility with constraints can be found in \cite{me-entcs08}.  Conformance (b) is a well-known {\em instanceOf} relationship between an object (instance) diagram and a class diagram (the metamodel) established over a typing mapping between the graphs such that the constraints are satisfied. We will discuss it in more detail in \sectref{isnt-metamod}

\begin{enumerate}[label={\em  Remark \arabic*)}, wide=0pt]
\item Workflow conformance includes not only control-flow conformance (\awmodok{the care of}{the primary approach in}{} standard process-based approaches to certification) but also data conformance, which makes it product-based as well.
\item In principle, the \refin\ chain can be extended with a deeper normative workflow \wfnormmeta\ regulating the definition of normative documents. 
\end{enumerate}

%
%
Thus, we reformulate assurance as
 \zdnewwok{a combination of (semi-)formalizable tasks, which are non-trivial\namodok{ and complex}{}{}, but known, partially well-studied and amenable to (semi-)automation.}{new pharsing}{-3em} 
{Using decomposition techniques described in
 \cite{me-models18}, 
 the conformance checks can be done in a hierarchical way, which would make them manageable even for very complex workflows.}




\newcommand\figinfer{\figref{fig:inference}}
\newcommand\figinf{\figref{fig:inference}}
\newcommand\figann{\figref{fig:GSNannotated-full}}
\newcommand\figanni{\figref{fig:GSNannotated-PD}}
\newcommand\figannii{\figref{fig:GSNannotated-PDD}}
\newcommand\figanniii{\figref{fig:GSNannotated-PDDR}}
\newcommand\figannlast{\figref{fig:GSNannotated-full}}
\newcommand\figtwotriangles{\figref{fig:two-triangles}}
\newcommand\figannlastlast{\figtwotriangles}
\newcommand\figmm{\figmetamod}
\newcommand\figrec{\figref{fig:gsn2wf}}
\newcommand\figgw{\figrec}
\newcommand\figgsn{\figref{fig:gsnEx-main}}
\newcommand\figrecsem{\figref{fig:gsn2wf}}
\newcommand\figgsntowf{\figref{fig:gsn2wf}}
\renewcommand\figrii{\figgsntowf}
\newcommand\figgsnann{\figref{fig:GSNannotated}}
\newcommand\figsepi{\figref{fig:metamodel}}
\newcommand\figmetamod{\figref{fig:metamodel}}
\newcommand\figsepii{\figref{fig:inference}}
\newcommand\figinst{\figref{fig:instance}}
\newcommand\sectsepdef{\sectref{GSNWF+}}

\section{From GSN to \wfp, 1: Process \& Data Flow}\label{sec:gsn2wf1}

\smallskip
\papertr{
\begin{figure*}[h]
\vspace{-1.3cm}
\centering
    \includegraphics[
                    width=0.95\textwidth
                 ]%
                 {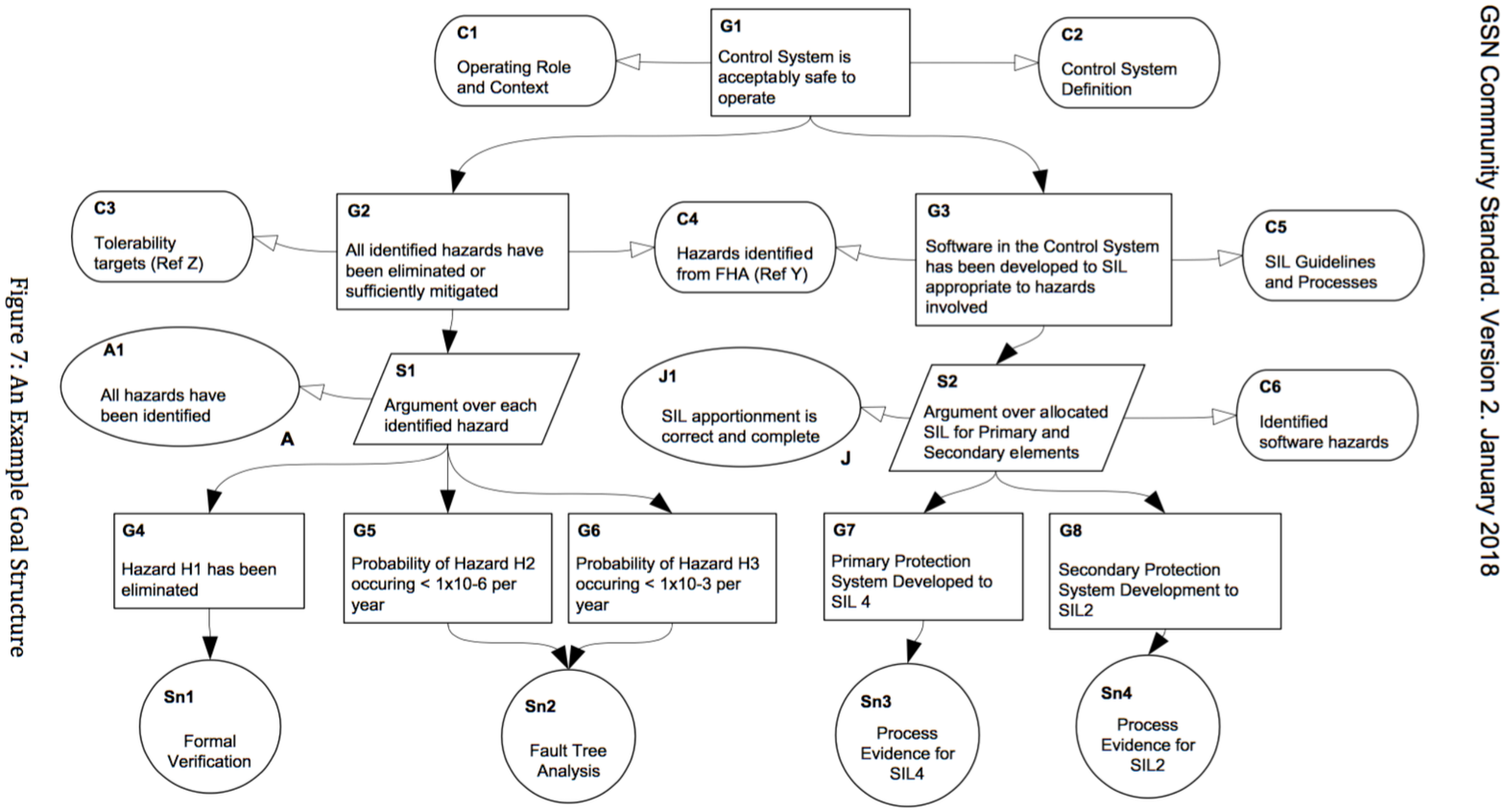}
                 \vspace{-1.5cm}
\caption{A safety case example from the GSN Community Standard \cite[p.14]{gsn18} \label{fig:gsnEx-main}}
\end{figure*}
}{
\begin{figure*}
\centering
    \includegraphics[
    				angle=90,
                   width=1\textwidth,
                 ]%
                 {figures/gsnEx-main}
\caption{A safety case example from the GSN Community Standard \cite[p.14]{gsn18} \label{fig:gsnEx-main}}
\end{figure*}
}
Figure~\ref{fig:gsnEx-main} shows a GSN goal structure taken from the GSN Community Standard \citegsNp{p.14}; we will refer to it as the (GSN) Example. We will interpret the Example as a sketch of a simple safety case, extract its semantic meaning and present it in \wfp\ terms, and then compare the two cases, as schematically shown in \figgsntowf. In more detail, we will extract from the data in the Example a SEP definition workflow \wfsep\ and show that the argument flow of the Example can be interpreted as checking conformance \paperGM{$\wfexe(X)\confcheck{\instan}{\VV}\wfsep$}%
{$\wfexe(X)\confcheck{\instan}{}\wfsep$} where $X$ denotes the Control System considered in the Example. \paperGM{We then somewhat artificially create a normative workflow to simulate the second conformance check.}{} Thus, all workflows \paperGM{occurring in formula \eqref{mainF}}{} are extracted from the data provided by the Example. 
\papertr{
\begin{figure}[h]
\centering
    \includegraphics[
                    width=0.45\textwidth%
                 ]%
                 {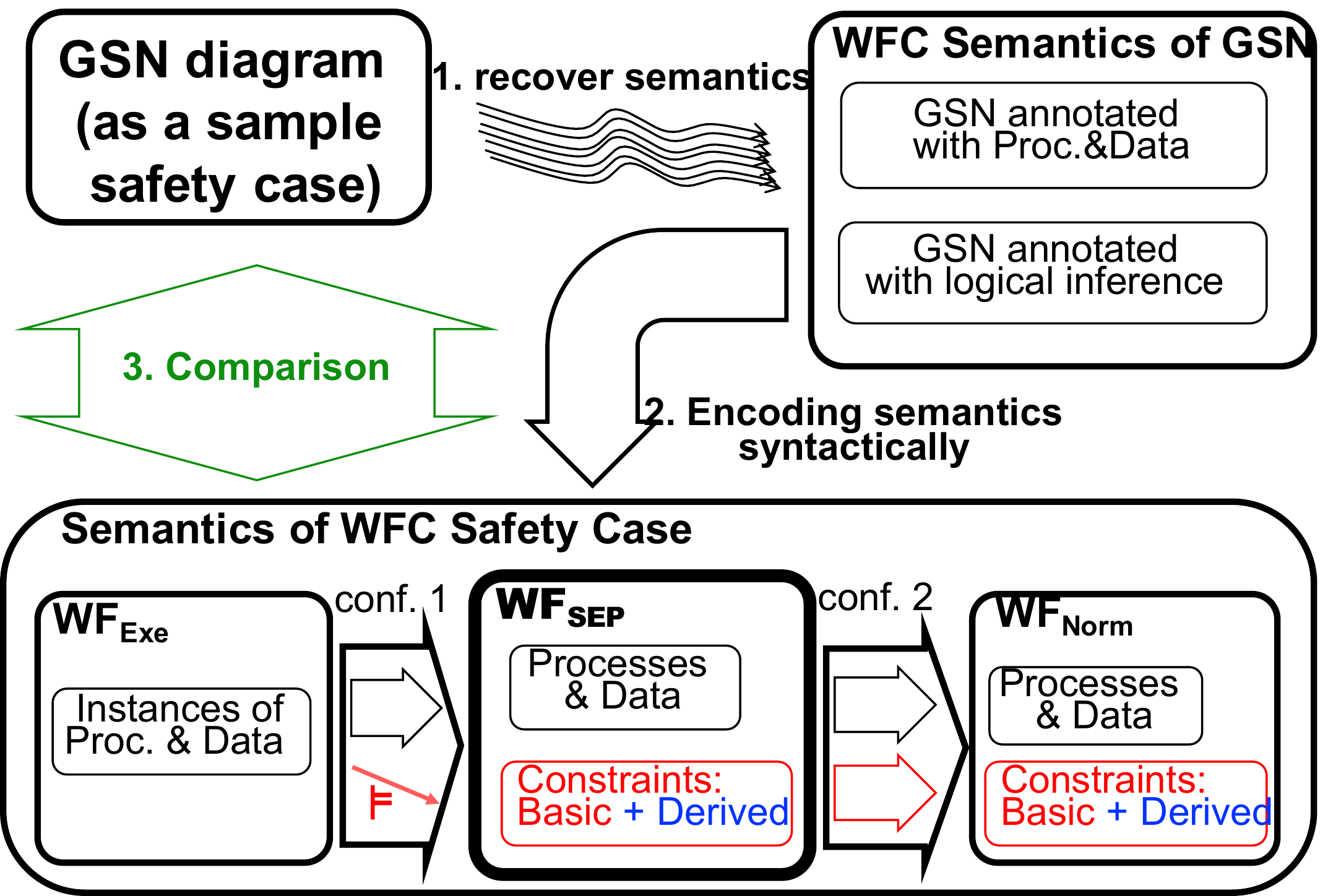}
\caption{From GSN to \wfp} \label{fig:gsn2wf}
\vspace{-0.5cm}
\end{figure}
}{
\begin{figure*}
\centering
    \includegraphics[
                    width=.5\textwidth%
                 ]%
                 {figures/recoverSem}
\caption{From GSN to \wfp} \label{fig:gsn2wf}
\end{figure*}
}

\subsection{Recover Semantics: GSN through workflow glasses}
\label{sec:recoverSem}

\newcommand\figtoinsert[1]{%
\papertr{\begin{figure*}
\centering
    \includegraphics[
    				angle=90,
                    width=1\textwidth%
                 ]%
                  {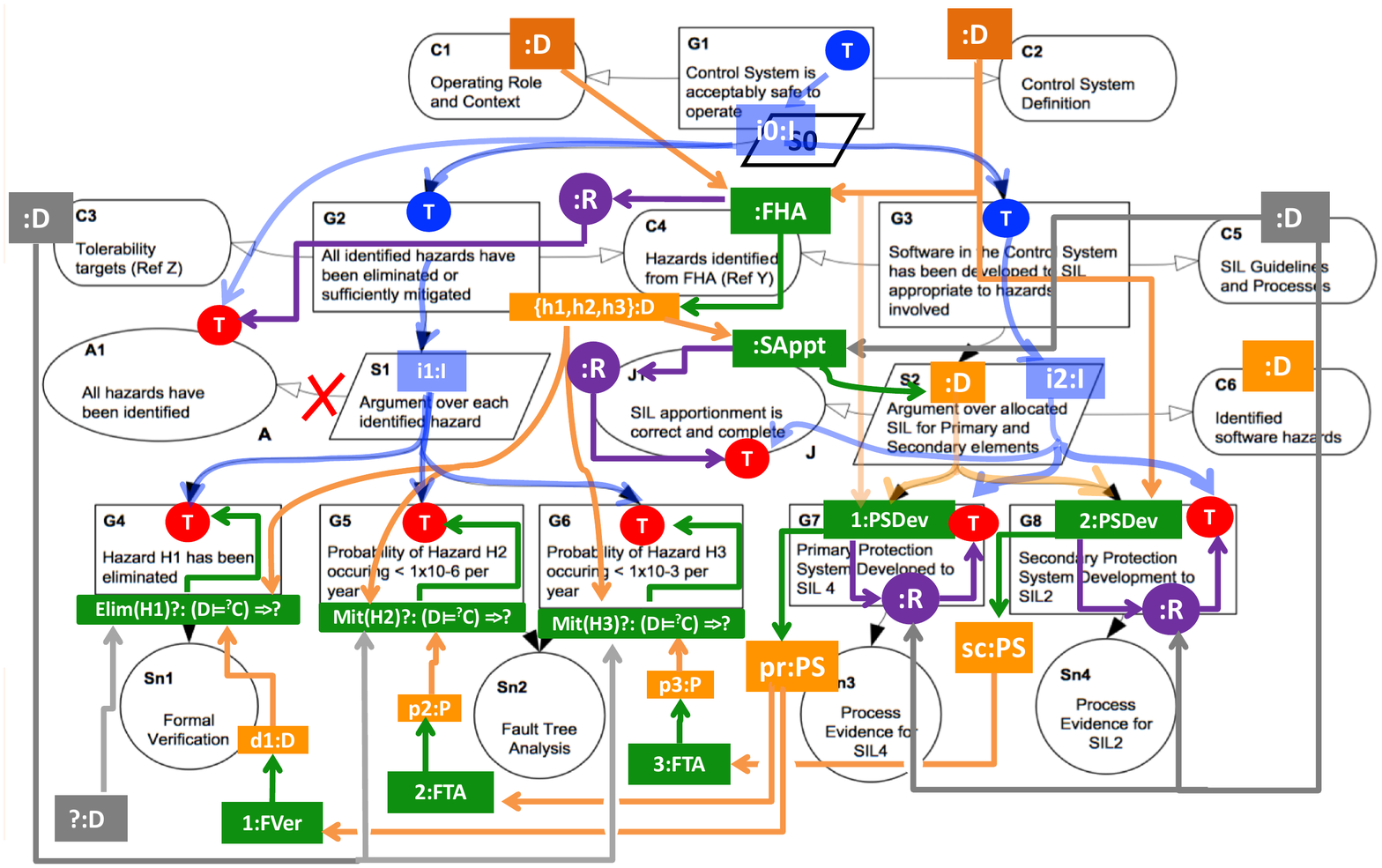}
\caption{GSN Example annotated with data (orange/grey), processes (green), review (violet) and inference (blue) blocks and dataflow arrows (orange/grey, green, violet and blue)
}\label{fig:GSNannotated-full}
\vspace{-.6cm}
\end{figure*}
}
{\paperGM{
		
	     }
	{\input{figures/TR/GSNannotated-#1}
	}
}}

Even a superficial inspection of the Example's elements  shows that they (directly or indirectly) refer to entities of several basic types. We will consecutively introduce them below.
\begin{figure*}
\centering
    \includegraphics[
    				angle=90,
                    width=1\textwidth%
                 ]%
                  {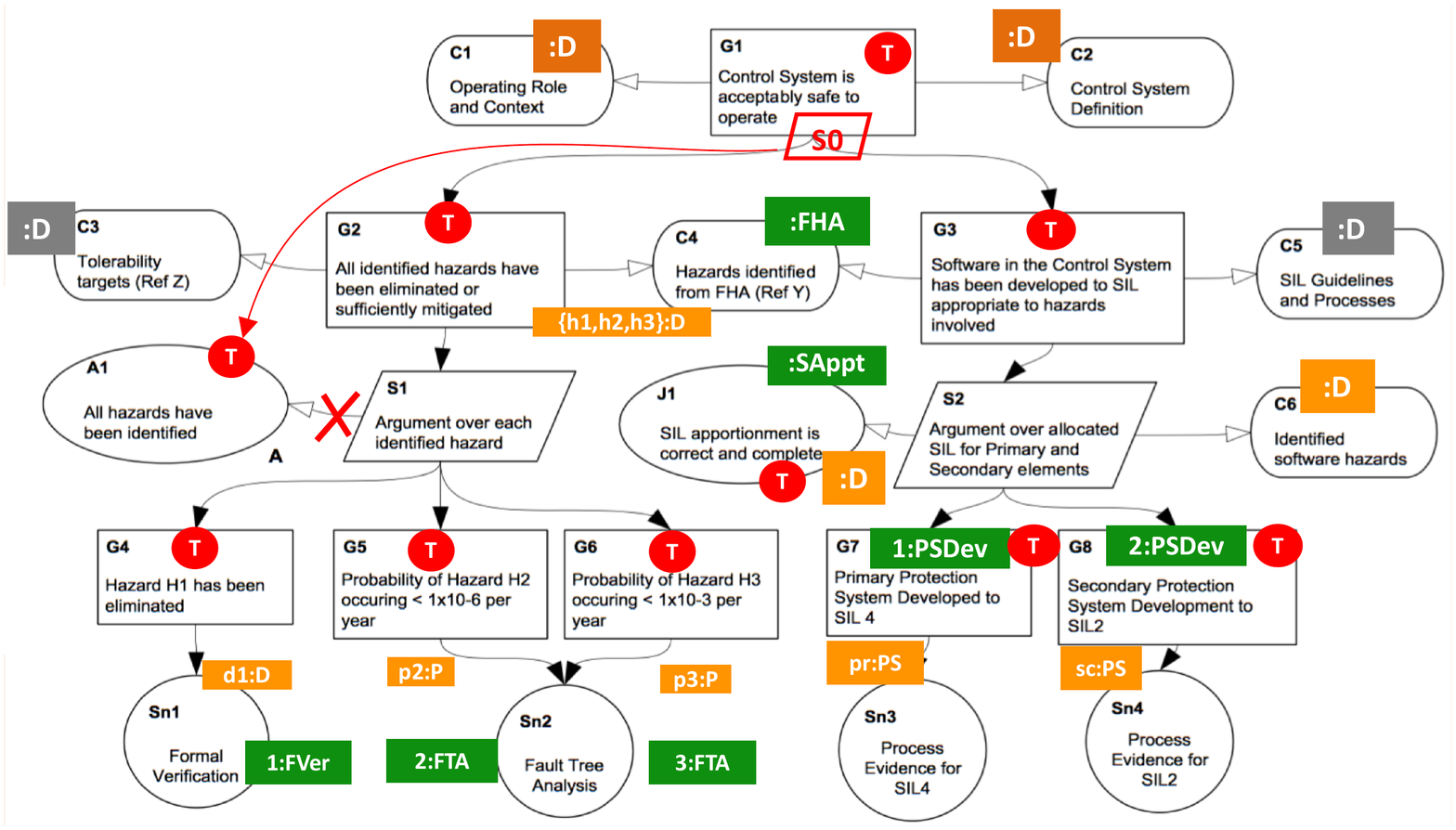}
\caption{GSN Example annotated with data (orange/grey) and processes (green) blocks.
}\label{fig:GSNannotated-PD}
\vspace{-.6cm}
\end{figure*}
%

\subsubsection{Processes and data.} Several elements of the GSN diagram explicitly refer to processes. 
For example, context C4 mentions Functional Hazard Analysis (FHA), justification J1 mentions SIL Apportionment (SAppt), solution node Sn2 mentions FTA (Fault Tree Analysis), and goals G7, G8 talk about Protection Systems Development (PSDev). 
There are also explicit references to data, \eg, context C4 says that FHA produces a set of hazards, and strategy S1 together with goals G4-6 say that this set consists of three hazards, H=\{H1, H2, H3\}, while goals G5-6 refer to probabilities of hazards H2-H3 occurrences.  

The results of such an inspection are shown in \figanni, where references to processes and data are labelled by green and orange block resp., and  special data residing in normative documents, Tolerability Targets in C3 and SIL Guidelines in C5, are shown in grey. (In addition, the input data for the entire SEP, \ie, the system data are shown with a darker orange shade.) A majority of data type names (such as Tolerability targets) are omitted and replaced by one generic name D to keep data blocks compact.  Moreover, logical data, \ie, truth values, are special and shown by red round blocks T, which should actually read T:Bool --- we again omitted the type to make blocks compact. 

Colons in block labels show that we talk about {\em instances} of processes and data, \ie, individual executions of processes and individual data objects consumed and produced by processes. 
For example, PSDev refers to a definition of some process of Protection System Development, which was executed twice (see goals G7, G8) for two hazard groups: one has SIL=4 and presumably consists of very bad hazards H1 (of SIL=4) and H2 (of high SIL 4 or 3), the other group consists of hazard H3 with SIL=2 (and hence the group SIL=2). Executing process PSDev for group 1 has resulted in a primary Protection System, pr:PS, and executing PSDev for group 2 produced a secondary Protection System, sc:PS; the systems are specified by different data of the same type PS. Similarly, we have two executions of the FTA process, which provide probabilities of hazards H2 and H3 resp., \ie, data p2 and p3 of the same type P(robability), \ie, reals in the segment [0,1]. 

 \namodno{Note a logical error in the Example:}{Note}{} assumption A1 is not needed for strategy S1, but is necessary for deriving G1 from G2 and G3 , \ie, it is a fundamental assumption of strategy S0 justifying this derivation. To fix this error, we added a strategy S0 and reassigned assumption A1 to S0 rather than to S1 (these corrections are shown by red). Another way to fix the error could be to reformulate goal G2 by erasing the adjective ``identified'' from its formulation.

\begin{figure*}
\centering
    \includegraphics[
    				angle=90,
                    width=1\textwidth%
                 ]%
                 {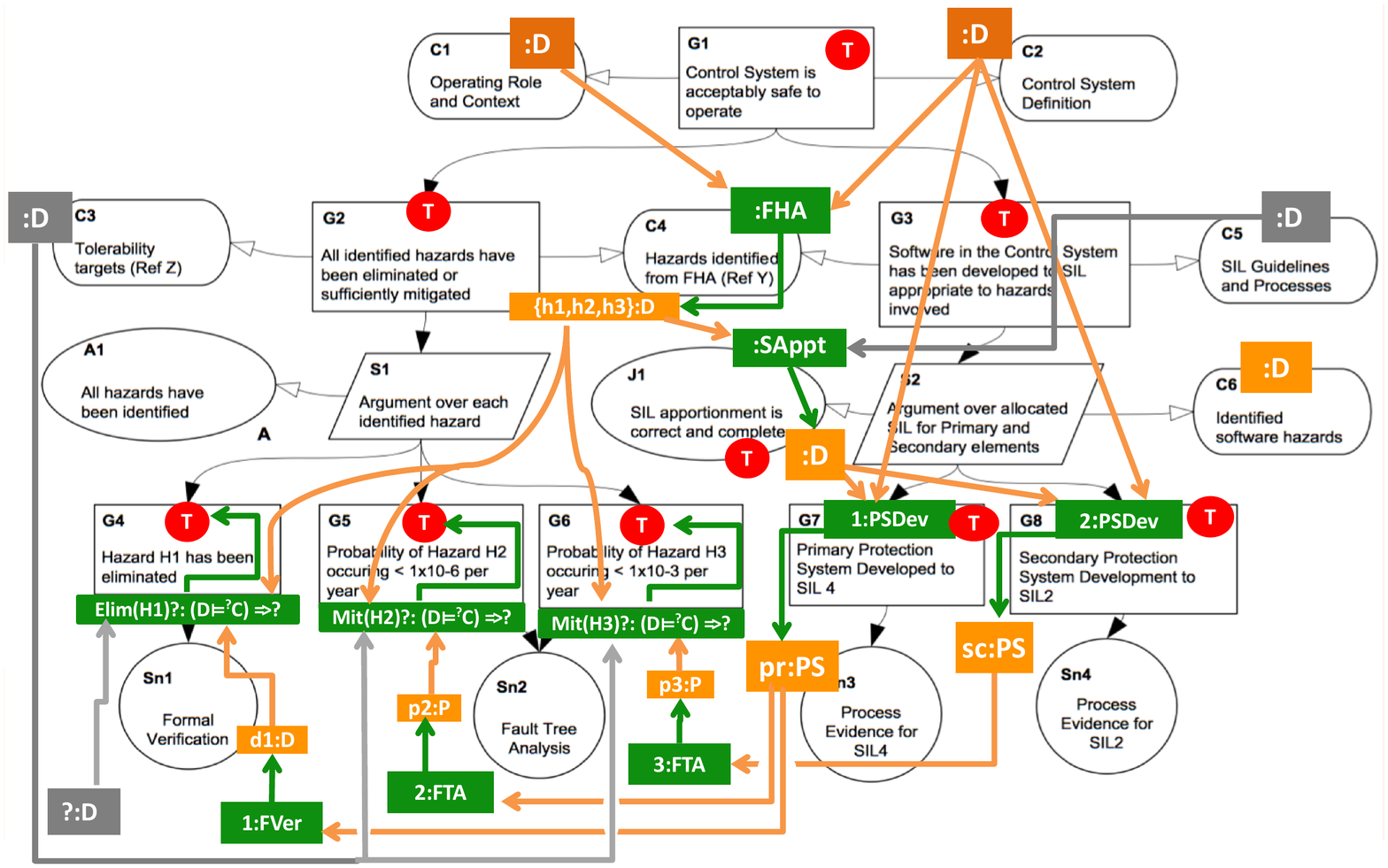}
\caption{GSN Example annotated with data (orange/grey) and processes (green) blocks and dataflow arrows (input---orange/grey, output---green).
}\label{fig:GSNannotated-PDD}
\vspace{-.6cm}
\end{figure*}

\subsubsection{Dataflow.}
The diagram in \figanni\ presents a very poor view of the story with all dataflow implicit. This is fixed in \figannii, where process input and output flow is shown by, resp., orange/grey and green arrows. The resulting diagram is self-explainable except, perhaps, the three processes below goals G4-G6 and their dataflow-- we will explain them now. 

Let's begin with execution 2:FTA. This process inputs data of the primary PS and performs FTA for hazard H2 providing its probability p2. Then this probability is compared with a certain normative value $10^{-6}$ (see goal G5 and the grey arrow showing the dataflow from the Tolerability Target document, TTD) and if $p2<10^{-6}$, we conclude that hazard H2 is sufficiently mitigated (formally, \Mit(H2)=\True). Note that while the comparison as such is a simple algebraic operation, the conclusion about the mitigation of hazard H2 is a much more complex process based on a nontrivial domain knowledge encoded in the TTD. To model such type of argument flow, we introduce a special type of processes $(D{\models^?}C)\implies ?$ that evaluates whether some data $D$ satisfy some constraints $C$ and what is a Boolean-valued consequence of such an evaluation. We thus assume that these processes consist of two parts: formal syntactical checking $D\models C$, and its safety-related semantic consequence.  Goals G5 and G6 encode two executions of such processes resulting in truth values \True\ for hazards H2 and H3 (note important dataflow links to these processes from hazards, which are assumed coupled with their SILs provided by FHA). We have a similar workflow for elimination of hazard H1: the formal verification procedure as such only provides data d1, while whether these data really show that hazard H1 has been eliminated is another issue addressed by the respective process \Elim(H1) as shown in the diagram. For uniformity, we added some normative data (note the grey block ?:D) intended to model the knowledge of how verification data to be interpreted to conclude that hazard H1 is eliminated indeed.

So far, we have shown the workflow producing the truth values in goals G4-G6, but other truth values are still isolated and their production is not explained.

\begin{figure*}
\centering
    \includegraphics[
    				angle=90,
                    width=0.9\textwidth%
                 ]%
                 {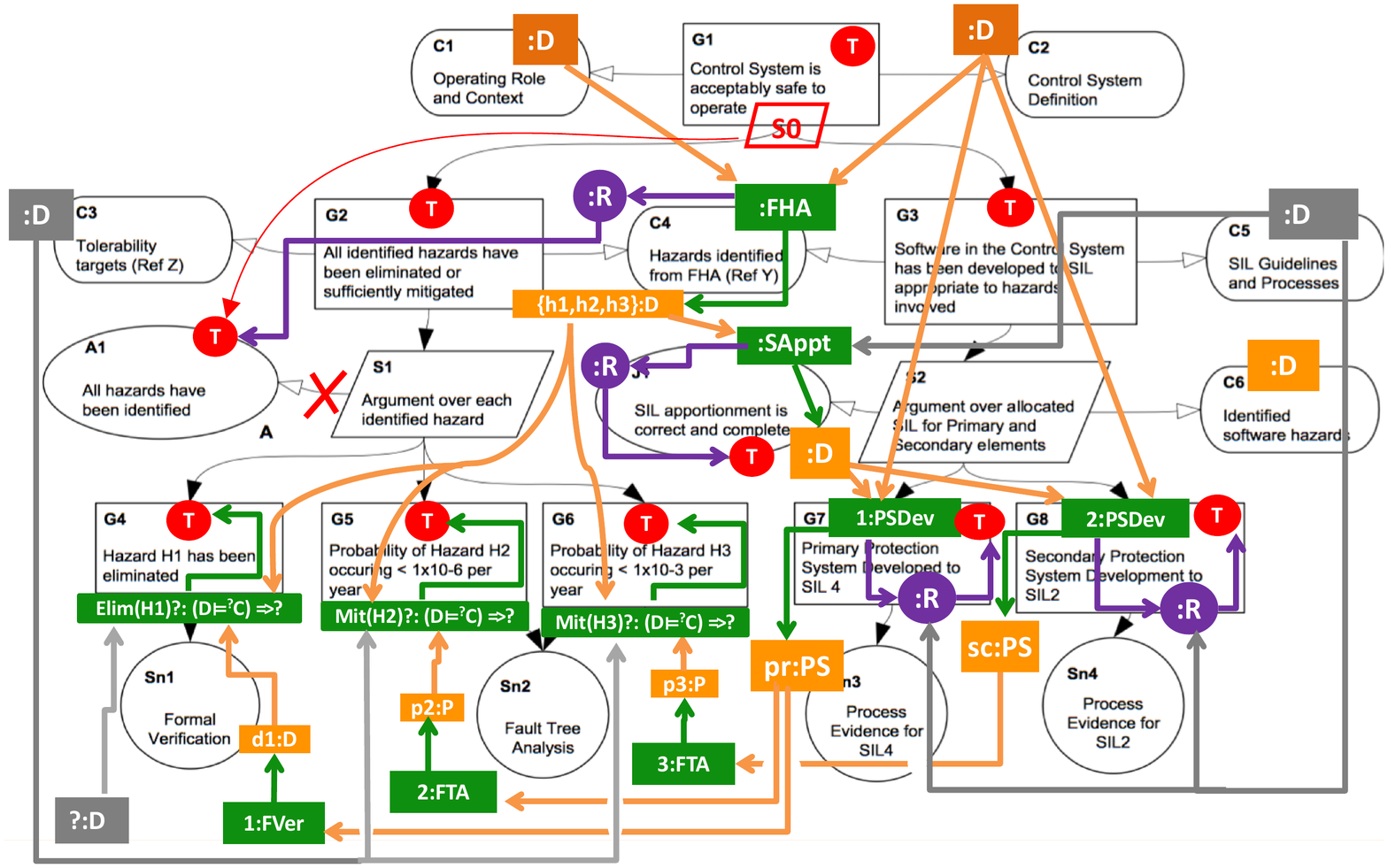}
\caption{GSN Example annotated with data (orange/grey), processes (green) and review (violet) blocks and dataflow arrows (orange/grey, green and violet)}
\label{fig:GSNannotated-PDDR}
\vspace{-.6cm}
\end{figure*}

\subsubsection{Reviewing.}
 The truth values in goals G7-8, and also in assumption A1 and justification J1  are produced by processes of a special type called {\em Review}.   Reviewing is  shown in \figanniii\ by purple round blocks labelled R. For example, to assure the statement required by justification J1, the process SAppt is to be reviewed, and similarly achieving goals G7 and G8 needs reviewing the \corring\ executions of PSDev. A review R of process P (we omit colons) takes P's execution data as its input  and outputs a Boolean value for the \corring\ proposition about P (typically, correctness and/or completeness phrased in context dependent terms, \eg, completeness of FHA as phrased in assumption A1).%
\footnote{To simplify notation, we connect P and R immediately, but with an accurate notation, there should be a data block in-between them. A detailed workflow model of reviewing will be considered later in \sectref{aspects}. }
We have thus explained all truth values besides the upper three assigned to the top goals G1-3.

\begin{figure*}
\centering
    \includegraphics[
    				angle=90,
                    width=0.9\textwidth%
                 ]%
                  {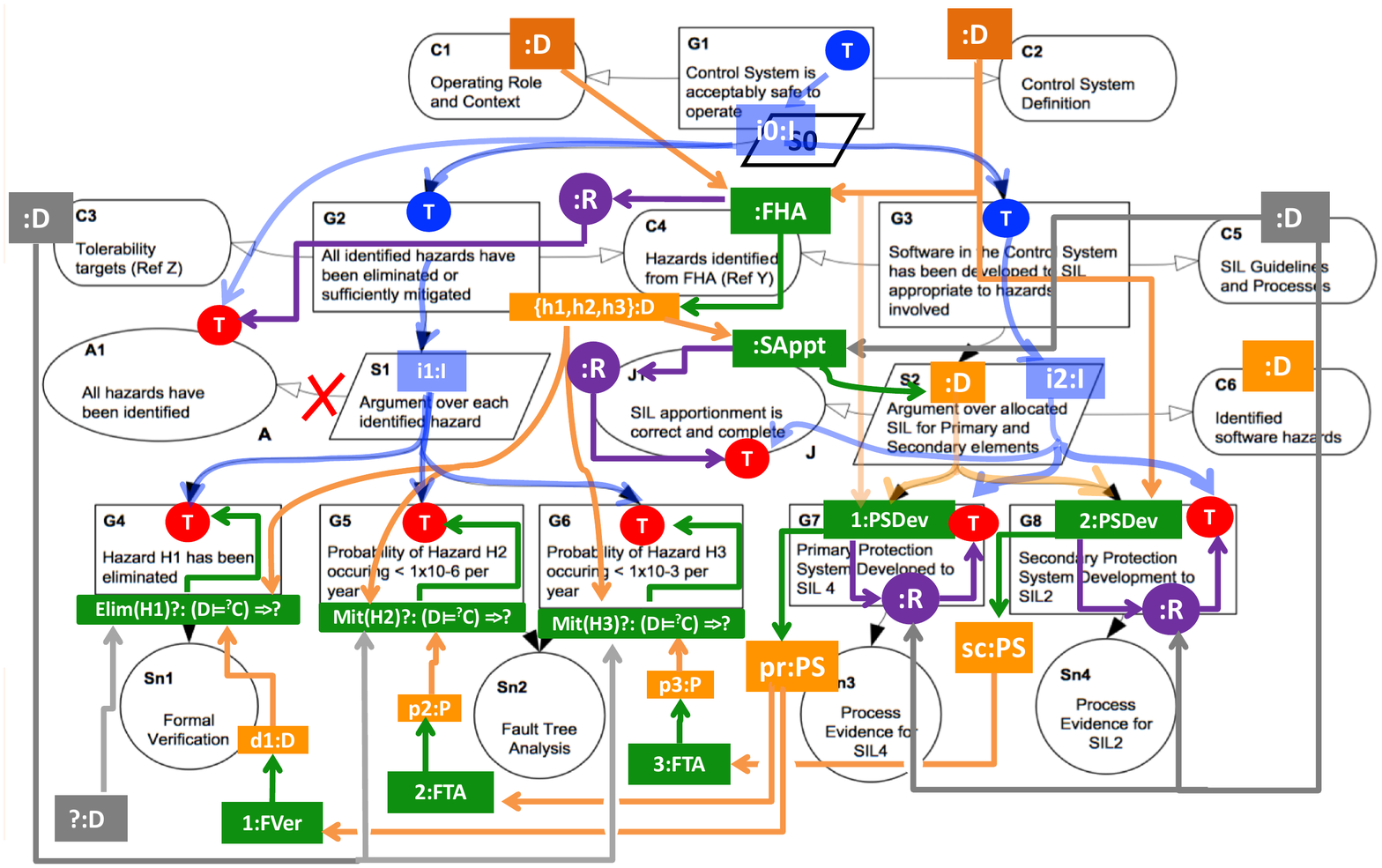}
\caption{GSN Example annotated with data (orange/grey), processes (green), review (violet) and inference (blue) blocks and dataflow arrows (orange/grey, green, violet and blue)
}\label{fig:GSNannotated-full}
\vspace{-.6cm}
\end{figure*}

\subsubsection{Logical inference.}
The last part of the story is logical inference: in \figannlast, note three light-blue blocks ix:I (x=1,2,0) residing in strategy nodes and light-blue arrows connecting them to propositions. The inference i0 (in strategy S0) can be specified by logical formula $A1\Rightarrow(G2 \,\& \,G3 \Rightarrow G1)$,  and similarly for the two other inferential blocks: inference i1 is unconditional while i2 requires a side condition J1. Although inference goes bottom-up, we show it with top-bottom arrows following the GSN notation to reduce clutter. 
We also show the truth-values produced by the inference flow rather than by the process-dataflow in blue. 

Note also that we changed the visualization of the dataflow from process :SAppt to proceses 1:PSDev and 2:PSDev to make it visually similar to the logical flow. We will discuss this later.   
   
In the next subsection, we will convert our process-data annotations into an accurate workflow model. 

\subsection{Instance Conformance: Syntax
} 
\label{sec:inst-conf-syn}

\papertr{
\begin{figure*}
\centering
    \includegraphics[
                    width=0.95\textwidth%
                 ]%
                 {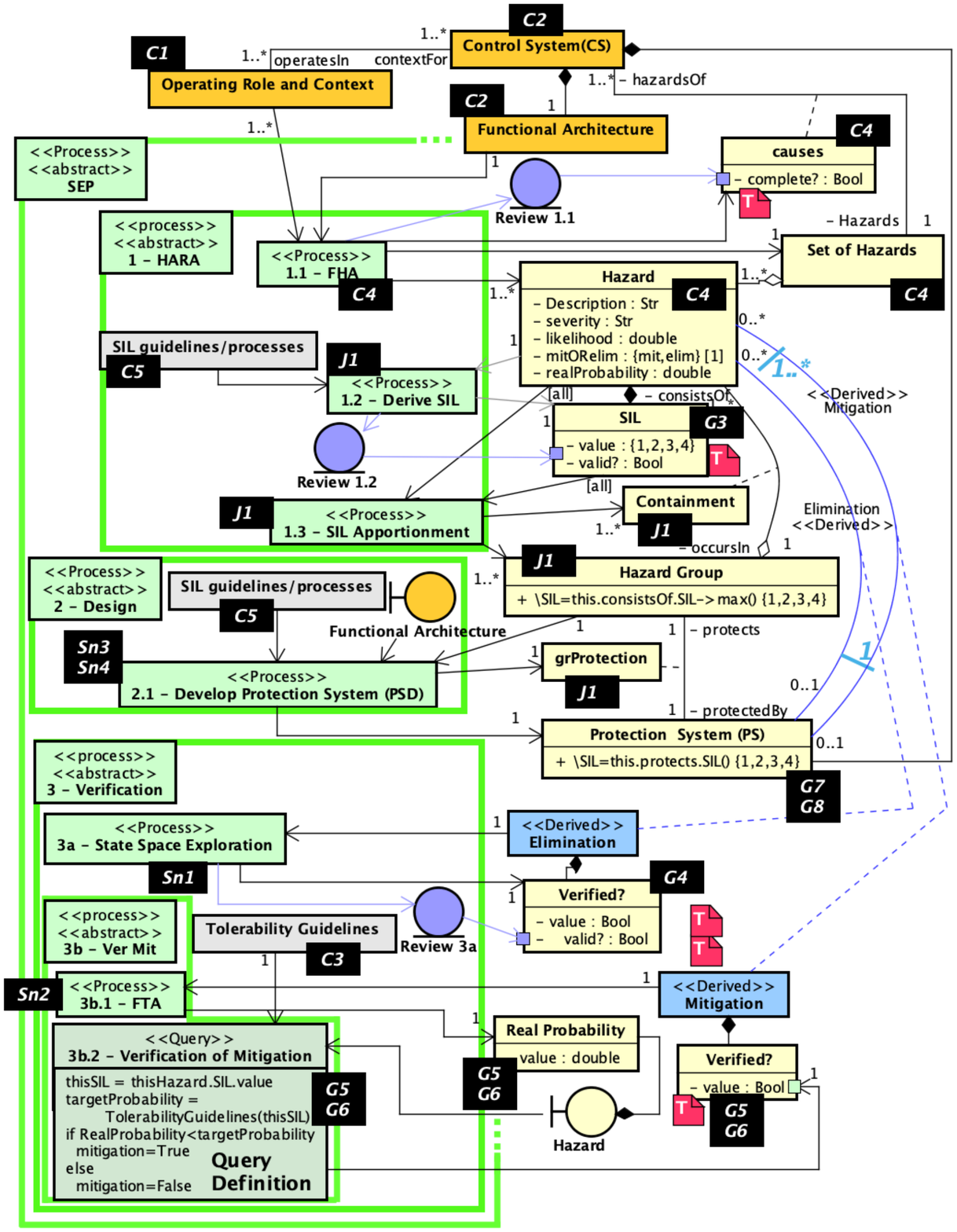}
                 \vspace{-0.5cm}
\caption{Metamodel further referred to as \wfsepdef\ 
} \label{fig:metamodel} 
\end{figure*}

}{
\begin{figure*}
\centering
    \includegraphics[
                    width=1\textwidth%
                 ]%
                 {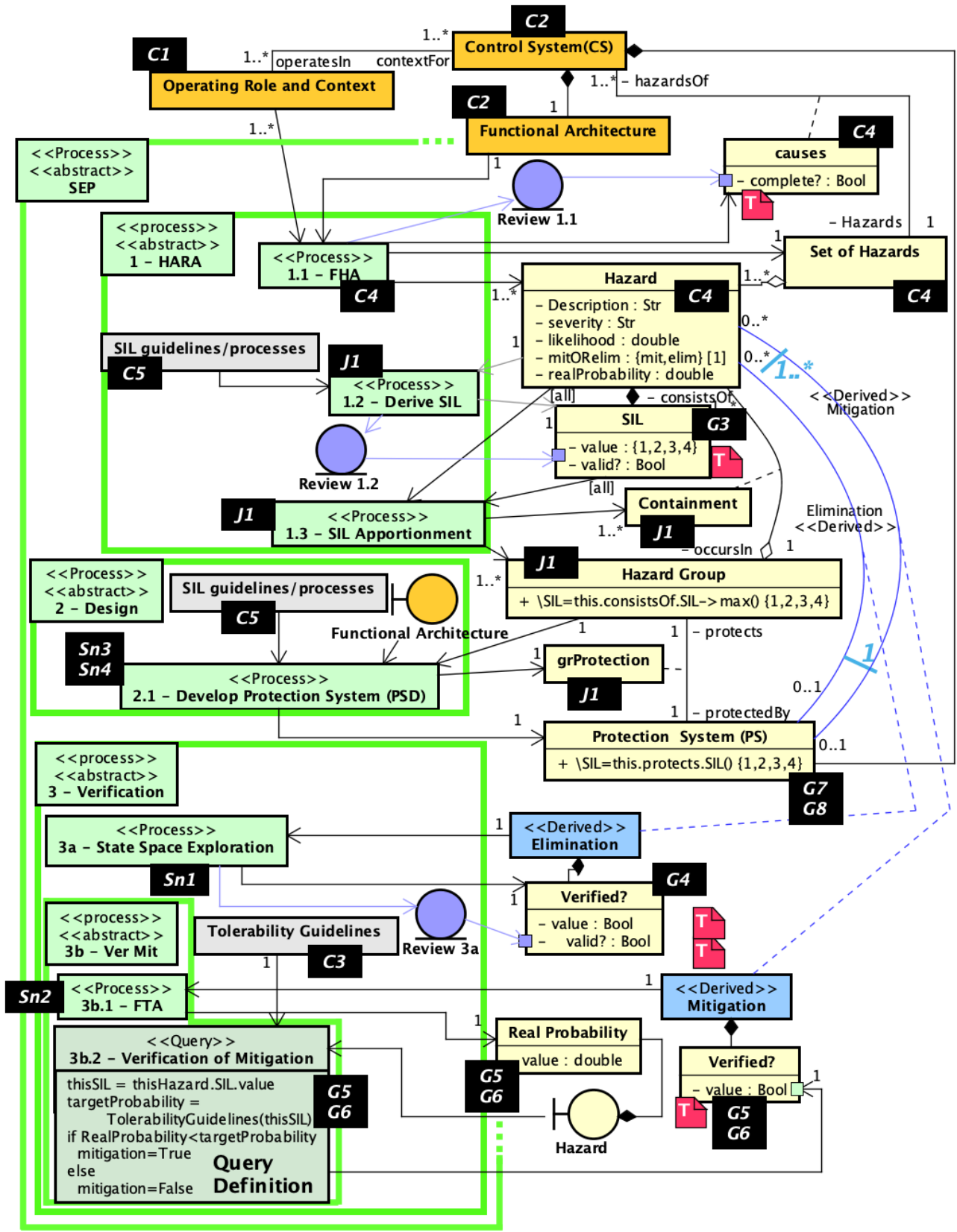}
\caption{Metamodel further referred to as \wfsepdef\ 
} \label{fig:metamodel} 
\vspace{-.7cm}
\end{figure*}

}
Process and data blocks in \figann\ sketch some workflow, which in this section we treat as an instance of some workflow definition/metamodel (\wfsep\ in term of \secref{assr-as-conf}). Making the latter syntactically accurate and semantically reasonable needs some domain knowledge beyond the workflow sketch in \figann, but we tried to \namodok{keep the extension beyond the Example story as minimal as possible.}{minimize the extension beyond the Example.}{}

We build the metamodel as a class diagram with a special type of class stereotyped Process (green in \figsepi), whose instances are process executions. Dataflow to and from processes is shown with directed associations, while ordinary (static) \asson s between data classes (yellow and orange) are all undirected and oriented vertically. Orange data are input to the entire SEP process, consisting \namoddok{from}{of}{}{0em} three components, HARA, Design, Verification (note three green rectangles in \figsepi), which are further decomposed as shown. The order of processes is shown with numbers prefixing process names (and can be inferred from the dataflow).
We use rectangle frames to show decomposition, \eg, process HARA consists of three subprocesses. 
Boundary objects (circle with a vertical line attached) are used as aliases for data to reduce the number of crossing lines in the model, and purple elements will be discussed later (ignore them for a while). \nanewok{Red notes are used to denote constraints on data. For example, the attribute valid? of SIL must be True. SysML-like ports are used on the sides of classes to indicate that an attribute is produced by a process. Attribute indentation denotes nesting (attributes of attributes)}{} 

\namodok{The diagram is mainly a self-explainable completion of the annotated Example (\figann); black-box labels show the backward mapping to the GSN diagram.}
{The diagram is mainly a self-explainable completion of the annotated Example (\figann). Black-box labels loosely show the backward mapping to the GSN diagram to illustrate how our metamodel was derived from it.}
{} 
We only mention several points. Note attribute mitORelim of class Hazard and its \multy\  [1], which says that each hazard is to be eliminated or mitigated but not both. If a hazard is to eliminated, then by composing \asson s \nmf{occursIn} and \nmf{protectedBy}, we obtain a derived \asson\ \nmf{Elimination} of \multy\ 0..1 as not each hazard is to be eliminated; similarly for hazards to be mitigated we obtain a derived \asson\ \nmf{Mitigation}.  Now \multy\ [1] implied \multy\ [1] spanning the two \asson\ ends (shown blue as it is derived)  says that each hazard is to eliminated or mitigated by exactly one PS. The inverse multiplicity 1..* is derived from \multi es of \nmf{protecs} and \nmf{consistsOf} and demands a protection system to protect against at least one hazard\nanewok{(this kind of association is not essential for safety, but is useful for ensuring there are no idle systems not protecting hazards, for example, after a design iteration)}{    maybe?}
Also, process FV in \figann\ is specialized as a state space exploration, which assures hazard elimination iff the hazardous state is not reachable. The comparisons residing in goals G5, G6 are specified as executions of the query specified in box 3b.2, which is defined to assure hazard mitigation.

Note a special feature of the workflow: data produced by a process are connected to its input data, \eg,\namodok{processes 1.2 and 1.3 take Hazard at the input and produce, resp., their  SIL attributes and (for process 1.3) HazardGroup together with as \asson\ linking it to Hazard. Or, p}{Process 3b.1 takes a pair (H,PS) consisting of a hazard H and a PS intended to mitigate H, uses this data to conduct a fault tree analysis, and outputs the Real Probability of H occuring with PS in place as an attribute of H.}{}

\papertr{
\begin{figure*}
\centering
    \includegraphics[
                    width=0.95\textwidth%
                 ]%
                 {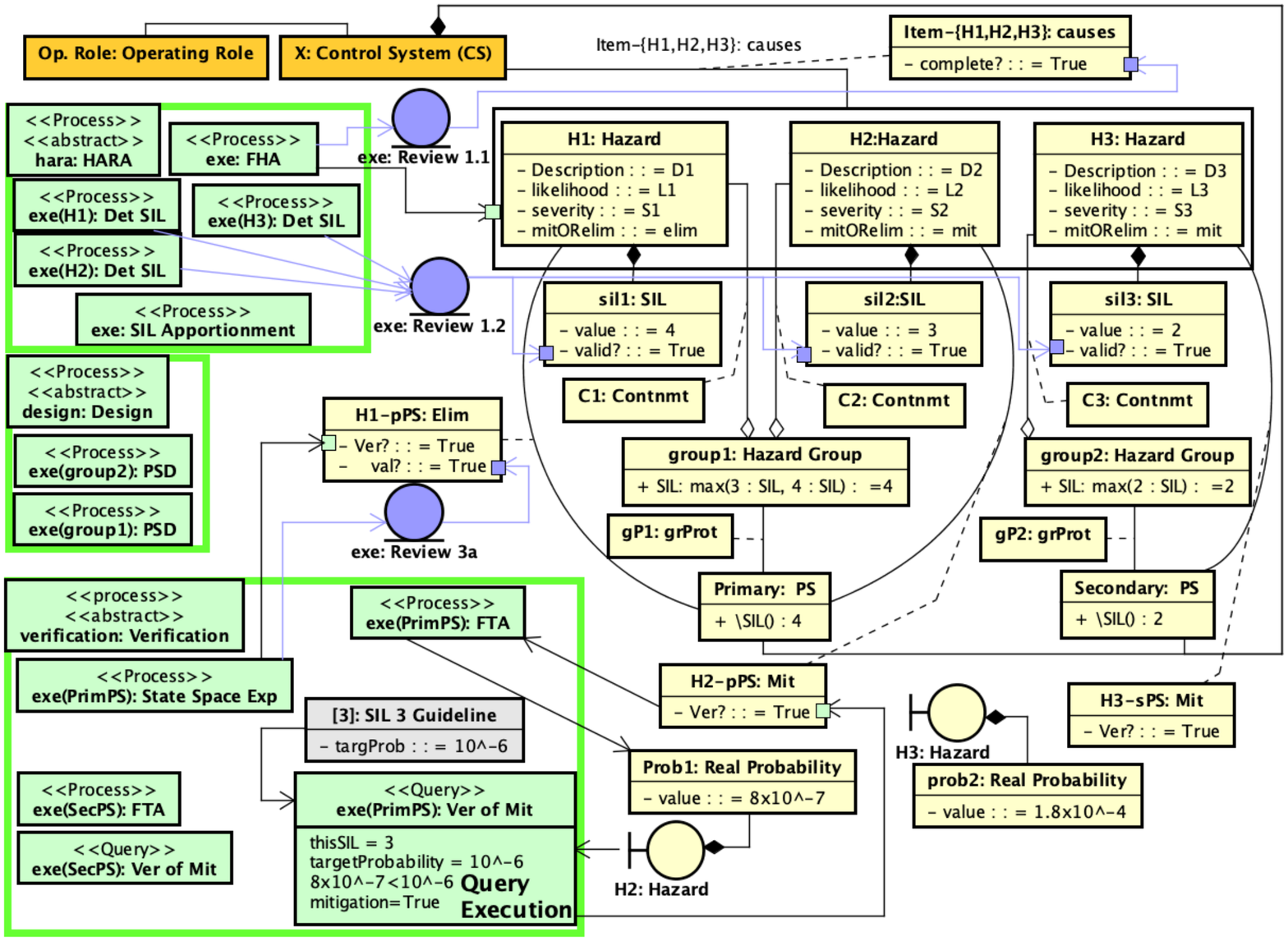}
                 \vspace{-0.7cm}
\caption{An \wfp-instance further referred as \wfinst } \label{fig:instance}
\end{figure*}

}{
\begin{figure*}
\centering
    \includegraphics[
                    width=1\textwidth%
                 ]%
                 {figures/instance.png}
\caption{An \wfp-instance further referred as \wfinst } \label{fig:instance}
\vspace{-.6cm}
\end{figure*}

}
The metamodel in \figmetamod\ defines a workflow, which can be executed for any data correctly instantiating the orange part of the diagram, and thus instantiate the yellow part of the class diagram \namoddok{(ISO 26262 refer to these as to WorkProducts)}{(commonly referred to as WorkProducts)}{seems random to mention 26262}{-4em}. The GSN Example actually tells us a story about such an instance \exeinst\ as shown in  \figinst, where the dataflow from (green) processes to (yellow) are omitted to avoid clutter (ignore purple underlined circles with arrows for a while) . It is easy to see that the instance workflow \exeinst\ satisfies all constraints declared in the metamodel workflow \wfsep.

\subsection{Instance conformance: Semantic validity}\label{sec:sem-val}
In assurance, there are two main questions about instances. The first is its syntactic correctness: whether all constraints declared in the metamodel such as \wfsep\ in \figmetamod\ are satisfied. This task can be done automatically with modern MDE tools. The second question is more challenging: syntactically correct data can be semantically invalid \eg, for instance \wfexex, a hazard can be missing from set {\sf Hazard.allInstances}=\{H1, H2, H3\}, or the SIL of hazard \namoddok{H1 is to be 2 rather than 4}{H3 is to be 4 rather than 2}{2 as a 4 is no big deal, but a 4 as a 2 is}{-4em} so that hazard grouping must be changed, or if even all SILs are valid, grouping was done incorrectly and hazard H2 is to be grouped with H3 rather than H1, etc. In fact, a majority of the instance elements are to be checked for semantic validity (we will say {\em validated}), including the results of verification. Validation of an element $x$ in instance \wfexex\ is done by reviewing the process that produced $x$, \eg, FHA for the hazard set completeness, SIL Determination for SILs validity \etc. Review processes (we say {\em reviews}) are denoted by purple underlined circles in the metamodel \figmetamod; we only show three of the total six (not seven as Query is computed automatically and does not need reviewing). Review executions in \figinst\ are also purple.  
 To simplify notation, we show the dataflow from a process to its review by an arrow and omit the intermediate data: this is just a syntactic sugar for a detailed specification considered in the next subsection.

\subsection{Reviewing as an aspect}
\label{sec:aspects}
\newcommand\figrR{\figref{fig:aspectR}}
\newcommand\citeR{\cite{?}\footnote{ref to paper[*]}}
\newcommand\citeRR{\cite{?}\footnote{ref to paper[**]}}
\papertr{
\begin{figure*}
\centering
    \includegraphics[
                    width=0.75\textwidth%
                 ]%
                 {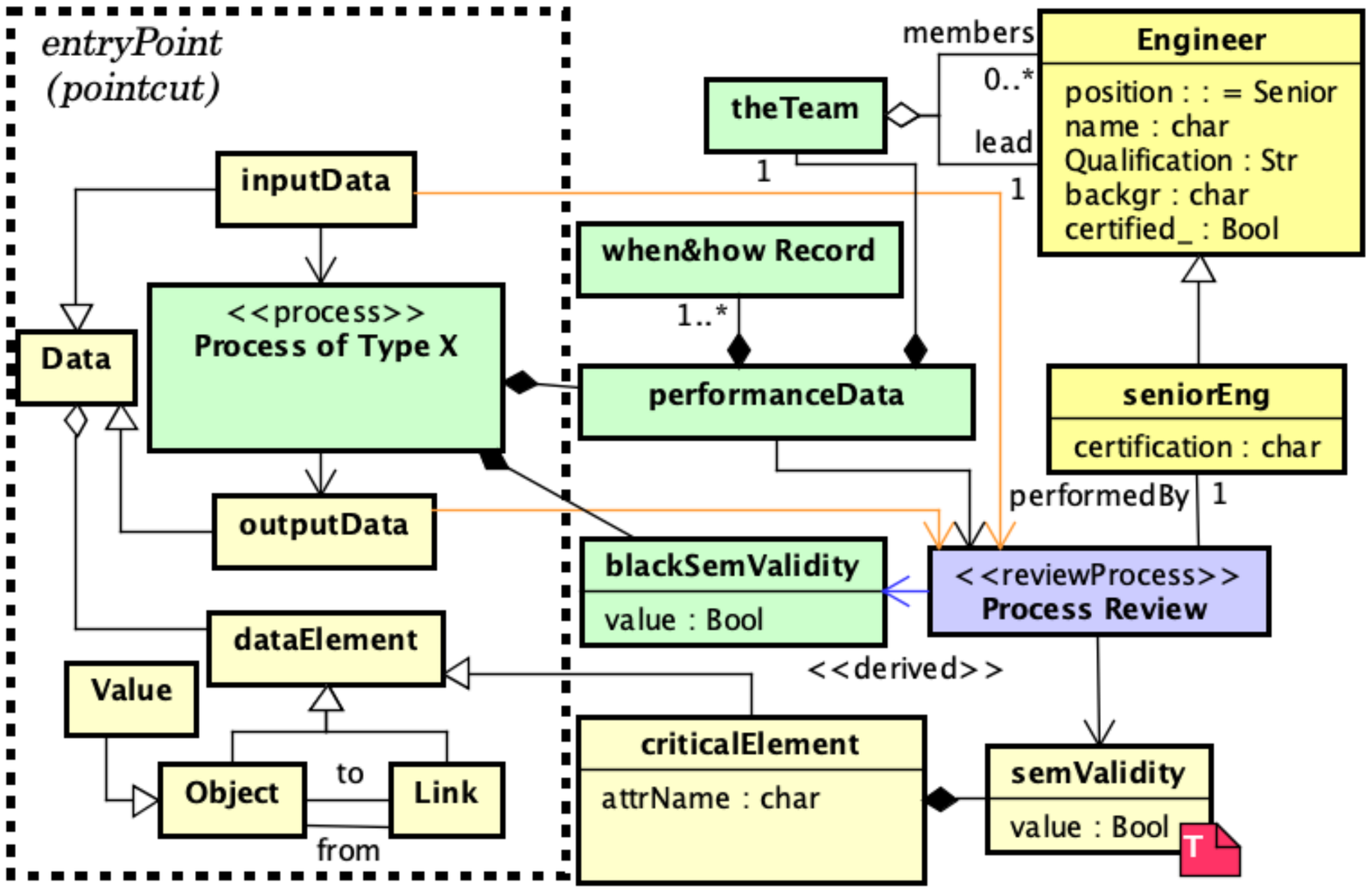}
\caption{Aspect Metamodel(Advice)} \label{fig:aspectR}
\vspace{-.6cm}
\end{figure*}

}{
\begin{figure*}
\centering
    \includegraphics[
                    width=1\textwidth%
                 ]%
                 {figures/aspectR-bbox.png}
\caption{Aspect Metamodel(Advice)} \label{fig:aspectR}
\vspace{-.6cm}
\end{figure*}

}
%


 \nanewok{The pattern in \figref{fig:instance} of a process producing data followed by a review checking the validity of that data is shown only once but in reality is repeated in many different places in a typical SC}{}. Thus, reviewing appears as a typical \awmodok{cross-curring}{crosscutting}{} concern and should be appropriately managed along the lines developed by the aspect-oriented programming community (AOSD), \cite{AspOrProg}. In our context, we need a mechanism of enriching a core metamodel with an aspect metamodel ({\em advice}, in AOSD terms) developed separately and weaved into the core metamodel at specified points  ({\em pointcuts}). We are not aware of any AO-\fwk s for metamodeling presented in the literature, and leave development of such for future work. Below we will use the main idea of aspect weaving \cite{advWeav} for metamodel weaving in the following sense. 

\begin{wrapfigure}{R}{0.45\columnwidth}
 \vspace{-0.5cm}
	\centering
\begin{diagram}
[w=5ex,h=5ex]
   \dbox{M_\main}&\lTo^{w_\enp} & \dbox{M_\entry} 
\\ 
	\dDashto<{e^*} & \mathsf{[PO]} 
	 &\dTo>{e}
\\  
	M_\main{+}M_\adv& \lDashto^{w^*} &\dbox{M_\adv}
\end{diagram}
\caption{Aspect weaving}\label{fig:aspWeaving}
\vspace{-2ex}
\end{wrapfigure}

A crosscutting concern is specified by an advice metamodel $M_\adv$, whose part $M_\entry\subset M_\adv$ specifies an entry, to which the advice is to be applied. Formally, we model this as an embedding mapping \incrar{e}{M_\entry}{M_\adv}. Now, having a main metamodel $M_\main$ with a set of entry points in it, for each such point \enp\ we specify a mapping \flar{w_\enp}{M_\main}{M_\entry} (see \figref{fig:aspWeaving}, in which given metamodels are framed and given mappings are solid, while computed mappings are dashed) and apply the pushout operation (PO) to the pair $(e,w_\enp)$, which will result in the merge $M_\main+M_\adv$ modulo the common (shared) part $M_\entry$: thus, we have weaved $M_\adv$ into $M_\main$ (details of how operation PO works can be found, \eg, in \cite{ehrig-book06}).   

In \figmetamod, three (from total six) entry points are shown with underlined  purple circles, and the advice metamodel to be weaved with them is described in \figref{fig:aspectR}.
It shows a very simple example of an actual review process around the key idea: the review observes performance data of a process, say $P$, including when and how it was executed and who the executor(s) were, alongside with the inputs and outputs of $P$. The review then outputs a Boolean value that (dis)approves the semantic validity of the critical data produced by $P$. \zdmoddok{(SEP designers can decide which information needs to be validated)}{What data are to be validated, and how reviewing is to be organized, is to be determined in the normative documents and be accordingly implemented in SEP.}{}{-0.5em}





\section{From \gsn\ to \wfp, 2: Argument Flow}\label{sec:gsn2wf2}

In \sectref{DDD}, we show how the data-driven inference mechanism of \wfp\ works We will try to follow the argument flow of the Example as much as possible, but be forced to deviate from it: the argument flow in the Example does not match its data flow, whereas our formal data-driven inference procedure forces the former flow to follow the latter. We  discuss the discrepancy in \sectref{anti-gsn} after we show how inference on the level of metamodel interacts with their instances in \sectref{isnt-metamod}.
{
\renewcommand\namod[3]{#2}
\subsection{Data-driven logical inference}
\label{sec:DDD}
\renewcommand\figinfer{\figref{xxx}}

\papertr{
\begin{figure*}
\centering
    \includegraphics[
                    width=0.95\textwidth%
                     ]%
                 {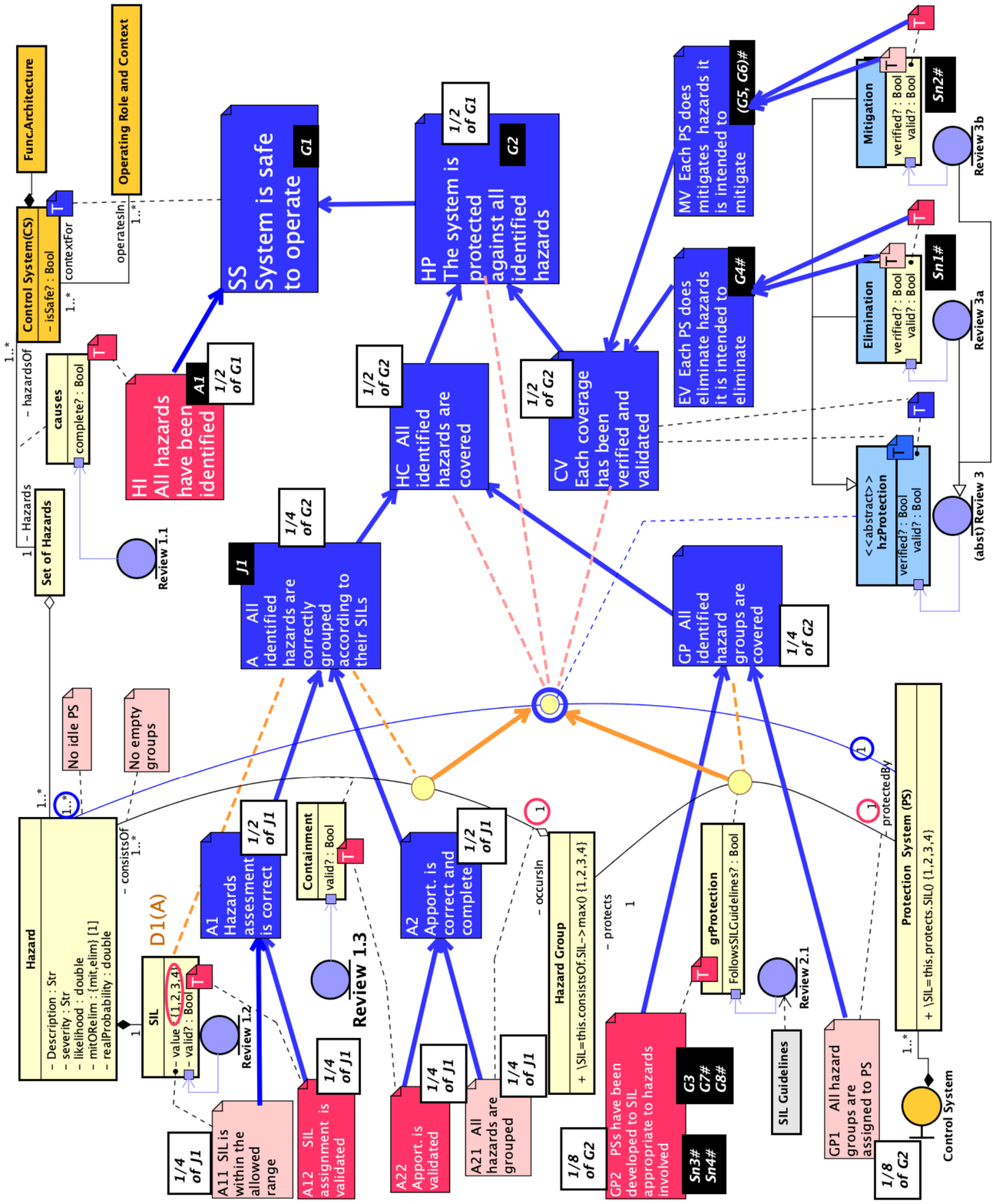}  
\caption{Argument flow(X\# refers to  the type for element X in \figref{fig:gsnEx-main})} 
\label{fig:inference}
\vspace{-.5cm}
\end{figure*}

}{
\begin{figure*}
\centering
    \includegraphics[
                    width=1\textwidth
                 ]%
                 {figures/inference}
\caption{Argument flow(X\# refers to  the type for element X in \figref{fig:gsnEx-main})} \label{fig:inference}
\vspace{-.6cm}
\end{figure*}

}
The \wfp-based inference is described by \figref{fig:inference}. It shows the metamodel \wfsep, in which green process blocks are omitted to save space (but implicitly are there) but aliases (in fact, entry points) of all six review processes are shown (in purple). There are several changes in the representation of the data part. We introduce an abstract reified association \hProtn\ (
generalizing associations \nmf{Elimination} and \nmf{Mitigation} (whose lines are omitted to avoid clutter), and derive from multiplicities for attribute mitORelim and associations \nmf{Containment} and \grProtn\ (see \figmetamod) that association \hProtn\ is composed 
\begin{equation}\label{eq:prt-def}
\hProtn = \Contt \comp \gProtn
\end{equation}
with multiplicities as shown (circled blue as they are derived). To shorten formulas, we will abbreviate the names above by \hpr, \ctt, and \gpr\ resp. 
The fact that association \hpr\ is composed from \ctt\ and \gpr, is shown in the diagram by two orange arrows from the component associations to the composed one
— this is a simple case of data derivation (or querying).
Important new elements in Fig. 7 are claims, i.e., natural-language formulations (in the assurance parlance) of data constraints, which are shown as notes attached to constraints. E.g., claims A11, A21 (pink notes on the left) are attached to encircled data constraints as shown, and claims A12, A22 (bold red and white font) are attached to validity constraints. 

Note an important difference between pink and red claims. Claim A11 is a simple syntactic constraint: it is violated if a hazard H does not have a SIL (all attributes have multiplicity 1 by default) or its SIL is an integer greater than 4. Claim A12 is a  semantic requirement of the validity of SIL assignment, e.g., if a hazard H is assigned SIL=2 whereas it is so dangerous that its SIL should be 4, then constraint A11 holds while claim A12 is violated. To formalize the latter as a constraint, we introduce SIL’s attribute \val? of type Boolean, produced by the corresponding review process, and require the value of this attribute to be T(rue). In GSN, conjunction of (syntactic) A11 and (semantic) A12 would typically be formulated in semantic terms as phrased in the claim note A1. 

In a similar way, we can treat claim A2 (below A1) as conjunction of a syntactic part A21 (the pink note) phrasing multiplicity 1 (note the dashed line), and a semantic part A22 validating whether the hazard grouping is done right w.r.t. safety engineering guidelines. Conjunction of A1 and A2 can be formulated as in the claim note A so that we have the inference tree from A11, A12, A22, A21 to A as shown in the diagram
We can identify claim A with context J1 of the Example, more accurately, constraint A can be seen as a formal model of (somewhat fuzzy) claim J1 (note the black box labelled J1 attached to note A). Based on this assumption, we somewhat arbitrarily distribute the meaning of J1 between its subclaims (note labels 1/4 of J1). 

Our seemingly pure logical work above was actually guided by data and dataflow in the metamodel. Two orange (the data colour) dashed lines from note A to class SIL (data element D1(A)) 
and association \Contt\ (element D2(A)) 
which point to data defining the validity of claim A. The meaning of A then guides us in finding the relevant data constraints A11 and A21, while processes that produce data D1(A), D2(A) point  to reviews needed for validation—the decomposition of A into conjunction $\bigwedge_{i≤2,j≤2}\mathrm{A}_{ij}$ is data driven! 


We can proceed in exactly the same way with decomposition of claim GP (below claim A that  refers to the Apportionment part of the story while GP stands for the GroupProtection part). We will mention briefly several points. The semantic claim GP2 is formulated in parallel to goal G2 in the example, and the name of the semantic validity attribute is changed from the common \valid? to a special name following the Example’s context C5. Then we infer 
\papertr{
	GP\implBy GP1 \& GP2 and HC\implBy A \& GP, 
}{
\begin{equation}\label{eq:derHP}
\mbox{GP\implBy GP1 \& GP2, HC\implBy A \& GP}, 
\end{equation}
}
where HC (HazardCoverage) is a claim about the composed association \hpr\ 
 (note the orange link). 
  We also know that the SEP requires to verify all hazard coverage links (which instantiate association \hpr). Verification processes evaluate attributes \ver? (whose Truth is required by corresponding pink constraints), which are then validated by reviews evaluating attributes \valid? (with red Truth constraints). We thus have an inference tree for claim CV as shown in the diagram. 
%
Next we assume that having a coverage link H-PS between hazard H and system PS (an instance of \hpr) verified and validated means that system PS does protect against H. We then derive  claim HP. Finally, assuming hazard analysis completeness HI, we derive the top safety claim SS as shown.   

 \subsection{Constraint derivation: From metamodels to instances}
 \label{sec:isnt-metamod}

 Constraints are elements of metamodels, and their derivation lives on the metamodel level while a typical GSN case builds an argument flow for a concrete system, \ie, an instance. The gap is bridged by the {\em soundness} of our derivation steps: we derive semantically rather than by syntactic rules.  
 In more detail, suppose we have a metamodel $M=(\gbf^M,\csetx{M}{\given})$ with \gbf\ being its graph of classes and \asson s, and \csetx{M}{\given}  a set of constraints defined over \gbf. An inference tree can be schematically presented by a sequence of inference steps
 \papertr{
 
  $ \csetx{M}{\given}=\csetx{0}{\der}\models_1\csetx{1}{\der}\models_{2}\ldots\models_{n}\csetx{n}{\der}=\csetx{M}{\der}\ni C!
 $}{
 \begin{equation}\label{eq:inf-seq}
 \csetx{M}{\given}=\csetx{0}{\der}\models_1\csetx{1}{\der}\models_{2}\ldots\models_{n}\csetx{n}{\der}=\csetx{M}{\der}\ni C!
 \end{equation}
}
 where  $\models_i$ is an inference step extending a set of derived \namodok{constants}{constraints}{} \csetx{i}{\der} with a new constraint $C_i$ so that $\csetx{i+1}{\der}=\csetx{i}{\der}\cup \{C_i\}$; the initial set consists of all constraints initially given in the metamodel;  and \csetx{M}{\der} is the final set containing the top claim $C!$. Soundness of a step $i$ means that for any instance $I$ properly typed over $\gbf^M$,  if $I\models\csetx{i}{\der}$ then $I\models\csetx{i+1}{\der}$. By transitivity, if $I\models\csetx{M}{\given}$ and all steps are sound, then $I\models C!$. %

Soundness of some of our derivation steps is based on simple logic, \eg, the inference 
CV\implBy EV \& MV is sound because of the data condition $\hPrtn=\nmf{Elimination}\cup\nmf{Mitigation}$, which is in turn implied by \multy\ 1 of Hazard's attribute elimORmit (see \figmetamod). Some inference steps can be seen as merely definitions, \eg, A1\implBy A11 \& A12. And some derivations are semantically non-trivial, \eg, when we conclude that having a coverage link H-PS both verified and reviewed implies that PS really protects against H. A common approach to such derivations is to require strong validation criteria so that getting equality \valid?=T for an instance $I$ is not easy but being achieved would indeed mean that the semantic condition in question is satisfied. 

Thus,  for a given system $X$, if $\wfexex\models\csetx{\wfsep}{\given}$, then $\wfexex\models C!$ so that the B) part of an \wfp-based safety case (see \defref{main}) consists of two major subparts: %
\label{page:b1b2}

\numitem{B1}  Constraint derivation over the SEP metamodel, $\csetx{\wfsep}{\given}\models C!$, 
  and 
  
  \numitem{B2} Checking instance conformance, $\wfexex\models \csetx{\wfsep}{\given}$, for a given system $X$, whose most challenging part is checking T for validity constraints (shown in red in \figinfer). 
}
\section{\wfp\ vs. \gsn: A Comparative Analysis}\label{sec:anti-gsn}

We will begin with a brief survey of how we processed the material described in the Example (further \exgsn) and come to its \wfp-version (\exwfp) to set a stage for the discussion, then present the results of \namodok{matching}{comparing}{} \exgsn\ and \exwfp, and conclude with some general observations about GSN. 

\subsection{From \exgsn\ to \exwfp: Defining the DCC}
We recognize that the main goal of the \exg\ diagram was to demonstrate the vocabulary of GSN elements and to show how they can usefully be put together \namodok{ to work for some purpose \exgoal}{}{} rather than write an SC per se. \namodok{However, for this \exgoal, they had chosen}{To do this, the authors of the document chose}{} to specify in GSN terms an oversimplified SC, which is a justifiable and reasonable choice: GSN is widely used for building and documenting SCs in many domains. 
The \exg-authors \namodok{did good work for}{skillfully managed to address}{} both goals: they described the notation and presented an interesting and seemingly well-arranged SC within a small \namodok{one-page}{}{} diagram.  This conclusion is justified by the fact that \exg\ has resided in the Standard without changes for eight years now, and  motivates\namodok{ and justify}{}{} our idea to use \exg\ as a sample SC to show how \wfp\ works\namodok{, and we processed it as schematically shown in \figrec.}{.}{} 

Our  comparative analysis needs an accurate description of what our processing brought to \exg. As \figann\ shows, \exg\ provides a listing of processes and data involved, and  \figmetamod\ shows that we did not intoduce in \exw\ new processes nor data classes (they all are labelled by the respective elements of \exg). However, classes (green and yellow) only give us a discrete view of the story without the dataflow between processes and data (green-yellow links), and \asson s  between classes (yellow-yellow); below we refer to both types of links as dataflow\zdmoddok{and call its two specializations dp-dataflow and dd-dateflow}{}{delete?}{-2em}. 
We have restored the missing structural links based on our understanding of the domain knowledge (DK) and thus came to metamodel \wfsep\ in \figmm. As this DK is sufficiently common, it seems reasonable to assume that the authors of \exg\ based their SC on a \namodok{very close}{similar}{} dataflow picture but kept it implicit. In addition, our \wfp-glasses revealed two more actors on the stage. The first is a set of constraints for the \df, and the second one is splitting the argument flow into two parts: (data-driven) constraint derivation over the metamodel, and checking the instance conformance $\wfexe(X)\confcheck{\instan}{}\,\wfsep$ (while a typical GSN SC considers argument flow for a given system $X$ without mentioning the two levels). 

We will refer to the differences between the two \fwk s, \GSN\ and \wfp, as to DCC (Dataflow, Constraints, Conformance). Again, it seems reasonable to assume that all three actors are implicitly (and perhaps inaccurately) present in the \exg\ so that the only real difference between the two \fwk s is that \wfp\ treats DCC as first-class citizens while \GSN\ keeps them underground. We can schematically write something like an equation 
\papertr{

\mbox{\hspace{1.5cm}\wfp  =  GSN + Explicit DCC}\hspace{1.5cm} (DCC)
}{
\begin{equation}\label{eq:DCC}
 \mbox{\wfp  =  \GSN\ + Explicit DCC}.  
 \end{equation}
}

 Our next goal is to compare the two cases, \exg\ and \exw, to see whether DCC-explication is practically important or is mainly a theoretical bonus.  

\subsection{Mapping between the two diagrams}
We compare two argument flow diagrams: \gsd\ in \figgsn\ (referred to as \exg\ or ``theirs'') and \wpd\ in \figinfer\ (\exw\ or ``ours''). In the latter diagram, note black labels that map elements of \exw\  
to elements of \exg. Some elements are a perfect match, \eg, our claims A and HP are exactly their justification J1 and goal G2 resp.; the labels can be seen as links specifying a \wtog-relation between the two diagrams.  \namodno{With this discipline, some of our claims have two labels, \eg, A has J1 and 1/4ofG2, where the fraction label refers to the place of the claim in the inference tree (fraction labels have other colouring schema to ease visualization), which makes relation \wtog\ of one-to-many type. }{With this discipline, some of our claims have two labels (fraction labels use different colours to ease visualization), which makes relation \wtog\ of one-to-many type. }{}

Finally, several elements in the lower part of our diagram are labelled X\# to be read ``the type of X'' where X is a GSN element: the latter may be an instance level datum, \eg, solutions Sn1 and Sn2 (as related to instance-level goals G4-G6) while our classes Elimination and Mitigation are meta-data to be labelled as Sn1\# and Sn2\# resp. 
Now some of the claims in our diagram have several labels, \eg, GP2 has label 1/8ofG2 by its place in our inference tree, label G3 as it is almost exactly \exg's G3, labels G7\#,G8\# because  conformance of an instance \wfexex\ to GP2 would correspond to \exg's goals G7, G8 and solutions Sn3, S4 resp. Having placed all labels and hence defined the relation \wtog, we can begin our comparative analysis. 

Mapping \wtog\ maps claims (nodes) in the \wpd\ to elements/nodes in the \gsd. For an ideal match between two diagrams, we would like this mapping to be compatible with the inference structure: if two our claims C1, C2 are mapped to GSN-elements E1, E2, and C1 is a subclaim of C2, then E1 should be a subclaim of E2 if they are both claims (this is an ideal case), or at least E1 is a ``subelement'' of E2 in some suitable sense, \eg, an assumption attached to a strategy S below goal G can be seen as a subelement of G. 
If the subelement relation is preserved but its depth is not, \eg, C1 is an immediate subclaim of C2 while E1 is a grandchild subelement of E2 (or the other way round), it would mean that one of the diagrams have a more detailed inference path than the other, which would still be a good match between the two inference trees. 

\begin{figure*}
\centering
    \includegraphics[
                    width=1\textwidth%
                 ]%
                 {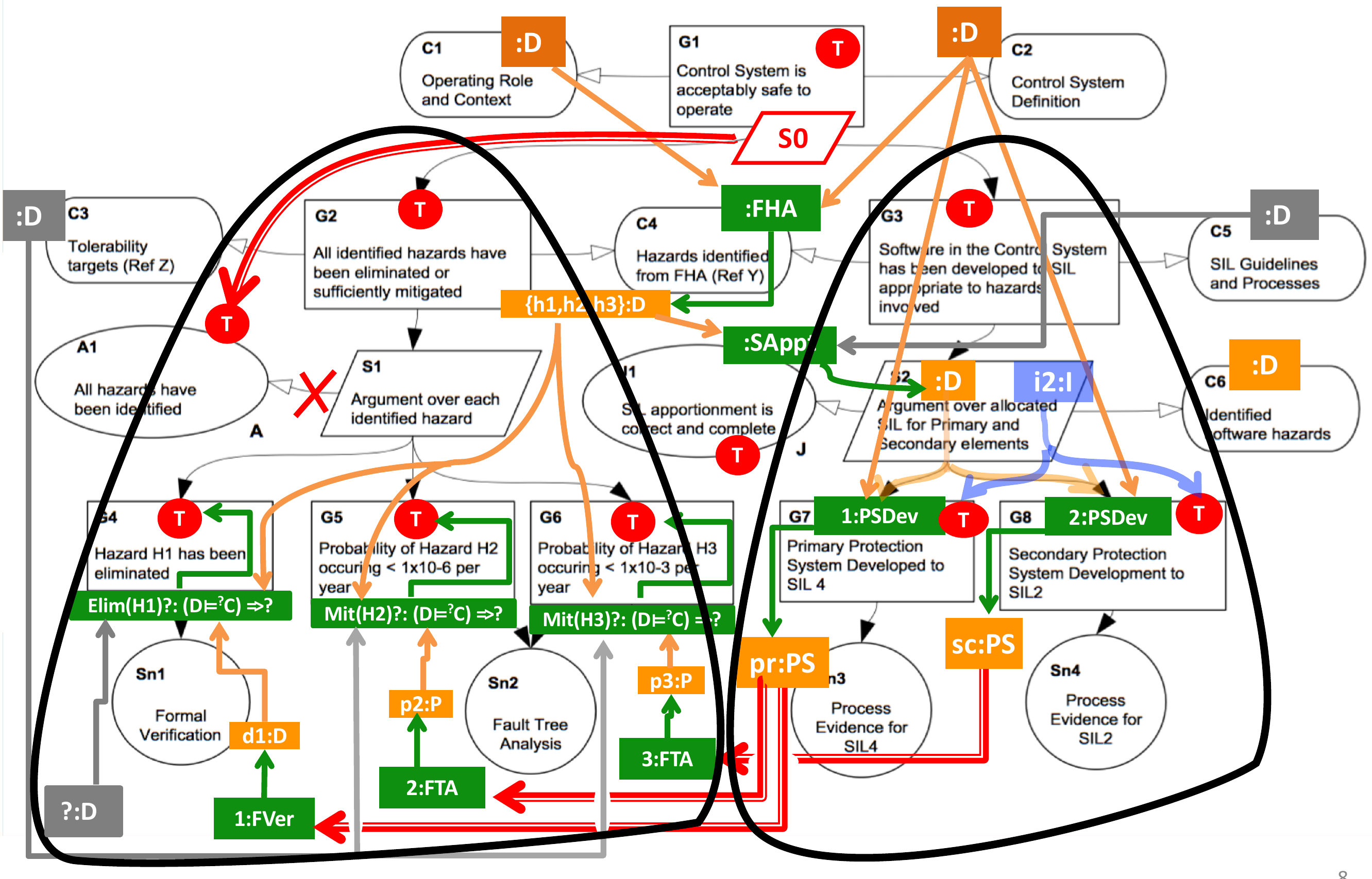}
\caption{GSN Example with some logical distortions explicated.
}\label{fig:two-triangles}
\vspace{-.6cm}
\end{figure*}

Surprisingly, the mapping \wtog\ between the diagrams  is not compatible with the inference structure and actually significantly distorts it. The most visible, and dramatic, distortion  is  the clear parallel structure of the \gsd\ which is not the case for the \wpd. In the \gsd, the flow below goal G1 splits into two parallel branches: below G2 and below G3, while in the \wpd,  the image of G3 is claim GP2 in the left lower part, and thus the G3-branch is embedded into the G2-branch.  Figure~\ref{fig:two-triangles} illustrates what happens: the dataflow specified by three bold elbow red arrows (at the bottom) actually map the right branch of the argument flow to the bottom of the left branch. 

Another distortion is the placement of assumption A1 as a subelement of G2 while actually it must be a subelement of the top  goal G1 in \gsd: see \figannlastlast, in which we introduced a strategy S0 to which A1 actually belongs. This displacement of A1 is logically obvious; it is surprising that it is going unnoticed for eight years.  \namodno{(An alternative way to fix the logical error could be to remove the attribute ``identified'' from goal G2 formulation.)}{}{}  
There are also smaller discrepancies such as having context C3 in the upper part of \gsd, while its proper place is  close to bottom as shown by \wpd. 

Another major issue is non-distinguishing of the instance-metamodel layers in GSN. E.g., in the \gsd, having a common Sn2 for two goals G5, G6 is correct on the level of types (see \wpd) but incorrect on the level of instances: it may happen that FTA was executed correctly for hazard H2 and incorrectly for H3, \namodok{then their reviews must be different and thus, on the instance level, we must have two execution of Sn2 as shown in our instance diagram \figinst.}{necessitating different reviews and thus two executions of Sn2 as shown in our instance diagram in \figinst.}{}   

\subsection{Comparison}

We will look at the mapping \GSN\ -- \wfp\ from the two opposite sides. 

\subsubsection{\wfp in the \GSN\ view.}The GSN literature identifies the two basic ingredients of safety assurance: {\em deduction} (the process of deriving new claims from the elementary claims until the top claim is reached) and {\em evidence} (that supports the elementary claims) \cite{RushbySRI2015} 
On the level of the \wfp-assurance architecture, we can identify GSN's deduction with the \wfp-task (\textbf{B1}) of constraint derivation (see the end of  \sectref{isnt-metamod}),
and GSN's evidence support with \wfp-task (\textbf{B2}): indeed, checking an instance conformance for a typical \wfsep\ basically amounts to checking non-empty instantiation of the type graph $\gbf^{\wfsep}$  that satisfies multiplicity constraints, and then checking that all validity attributes have value T, which is close to providing evidence in GSN. As for elementary GSN blocks from which GSN diagrams are built (such as Goals, Assumptions, etc.), \figannlast\ on p.\pageref{fig:GSNannotated-full} shows that every element of the GSN vocabulary gets its representation in \wfp. Thus, \wfp\ is expressive enough to model GSN on both levels: the architecture and the atomic blocks. 

\subsubsection{\GSN\ in the \wfp\ view.}

We will begin with several technical observations (items a), b) below), and then discuss their general consequences in items c-g). 
 \begin{enumerate}[label={\em  \alph*)}, 
	wide=0pt]

\item Although a GSN diagram can include processes and data (see \figanni\ on p.\pageref{fig:GSNannotated-PD}), it does not have primitives to model dataflow (\figannii). The result is that dataflow is lost (eg, as shown by \figtwotriangles\ on p.\pageref{fig:two-triangles}) or is (mistakenly) modelled by the argument flow: note the parallel structure of the dataflow and logical flow over strategy S2 in \figannlast\ on p.\pageref{fig:GSNannotated-full}, which may provoke errors. 

\item GSN flattens the two-level structure of checking the SEP execution  on the elementary level (ref to the FTA thing), and on the conceptual/architectural level of constraint derivation {\bf (B1)}  and instance conformance {\bf (B2)}.

\item A serious consequence of the lost dataflow is the absence of explicit traceability mechanisms (cf. discussion in \sectref{compara-survey}).

\item Flattening (item b) often hampers comprehensibility and may lead to errors, which appeared even in our simple case (the same solution for two hazards) and can dangerously accumulate in large industrial safety cases. 	
	
\item The GSN approach to structuring seems ad hoc. In \wfp, the inference tree is built from claims (constraints) while GSN's inference involves several types of elements (note that claims in \figinfer\ are labelled by G, A, J in a seemingly unordered way \namodok{types of GSN elements}{}{}). Labelling in \figmetamod\ shows a close relationship between data and context, but other types also appear there. 

\item While in \wfp, graph-based object-oriented data drive the inference and thus facilitate building the safety case, GSN leaves safety case builders on their own with the problems of proof and evidence design, which can lead to an ad hoc inference structure as we have seen above. 

\item Overall, having the DCC trio (Data-Constraints-Conformance) explicit and formalized is a way better for building and documenting safety cases (cf. discussion in \sectref{compara-survey}). 
\end{enumerate}

\section{Assurance as conformance, revisited: Parts and colours of safety assurance}
\label{sec:assr-as-conf-rev}

We first argue that the CAE interpretation of the GSN diagrams can be very much misleading. Then we recall the results of the previous section 
to show how the major question of safety assurance is answered in the \wfp-fwk, and why we need the second conformance check $\wfsepdef\confcheck{\mathsf{ref}}{}\wfnorm$. The last section provides some details.  

\subsection{GSN vs. CAE}
GSN-inspired but \wfp-realized safety case would consist of three conceptually different parts. 

{\bf Part 1.} The SEP definition workflow \wfsepdef, is to be provided. It includes all process and dataflow (green and orange), including reviewing (violet) and, importantly, constraints (red) over data and dataflow. 

{\bf Part 1*.} From constraints given in Part 1, new constraints/claims are derived (blue), and the corresponding inference tree eventually results in the top safety claim. 

{\bf Part 2.} The SEP execution workflow for system $X$ is to be provided, $\wfsepexe(X)$, and demonstrated that the execution conforms to the definition, $\wfsepexe(X)\confcheck{\instan}{}\wfsepdef$. Specifically, the execution instance has to satisfy all given (red) constraints (which GSN would call ``the body of evidence''). Then, by the transitive soundness of inference (see \secdercon), we conclude that the instance satisfies the top safety claim too. 

Importantly, the blue inference can be seen as a ``proof'' of the safety claim if only each inference step is semantically sound (\secdercon)--- this fundamental aspect of the argument flow is often underestimated in the GSN literature and, we believe, in the practice of writing safety cases in the GSN notation. Indeed, assuring soundness requires a justified strategy assigned to each decomposition/inference step, \ie, a strategy and its justification at each decomposition step are to be mandatory elements of GSN-based safety cases. However, the GSN syntax does not make a justified strategy a mandatory element, and many safety cases appearing in practice do use this liberty, \eg, the GSN Community example we considered above. 

Moreover, a majority of GSN decomposition steps are not of logical nature and show a process decomposition rather than logical inference. For example, suppose we want to provide a more detailed evidence for the claim  {\em HI: All hazards have been identified} (see \figinf\ on p.\pageref{fig:inference}) rather than just FHA reviewing. Then we need to decompose FHA into smaller subprocesses, then decompose the latter into smaller subroutines and so on, and \corring ly, FHA reviewing will be composed from more elementary reviewing acts. The result will be a refined SEP definition, over which we can build a refined inference tree, but the latter is actually not needed: as we have seen in  \sectref{DDD}, inference is a straightforward (and practically trivial) consequence of the process and data decomposition. The essence of deriving claim HI is in the FHA decomposition, and each step of such decomposition is to be justified and assured to be semantically sound in the following sense: if all subprocesses $P_i$ of a process $P$ are well done, then the process $P$ is well done. 

The decomposition we are discussing can well be phrased in terms of goal decomposition: To assure that the system is safe (SS or G1, see \figinf), identify all hazards (HI or A1) and protect the system against each hazard (HP or G2). To identify all hazards (HI), do the FHA in the following way .... (see the above). To protect the system against each hazard (HP), make sure that all identified hazards are covered (HC), and each {\em beingCovered} relationship is verified and validated (CV). To ensure each hazard has been covered, make sure that apportionment is done well (A) and all hazard groups are covered (GP), and so on. Thus, what is shown in \figinf\ as inference, is actually nothing but goal decomposition or, technically, a SEP decomposition into FHA and HP (hazard protection subbranch), followed by HP's further decomposition into HC and CV, and so on as shown in \figinf\ or in more detail in \figmetamod\ (p.\pageref{fig:metamodel}).  Phrasing this decomposition in logical terms (CAE) can be misleading as it creates an impression of a ``proof'' that decomposition is done right (\figinf), while actually such a ``proof'' is missing from \figinf. 

Indeed, how do we know that {\em if i) all hazards are identified and ii) the system is protected against each identified hazard, then the system is safe}? In the logical sense, this is just a tautology: if a hazard is something that can make a system unsafe, then the system is, by definition, safe iff it is protected against all hazards. Other inference steps in \figinf\ are also mainly tautological if considered in a purely logical way because they contain not only syntactic verification (which is formal and thus indeed logically inferential), but also semantic validation part. For example, the syntactic part of claim HC (\ie, the multiplicity 1 of \asson\ 'protectedBy') can be formally derived from the syntactic parts of claims A and GP (which are multiplicites 1 of their \corring\ \asson s traced by dashed ornage lines) based on the equality (\eqref{prt-def}) on p.\pageref{eq:prt-def}---this is a logical inference step, but it is very simple. As for the semantic part, we need either to postulate a definition that HC is A \& GP (which makes the inference HC \implBy\ A \& GP trivial), or to consider HC as being semantically different from A \& GP and thus making the inference step HC \implBy\ A \& GP semantically non-trivial and hence requiring a justification.  The latter would refer to previous experience and expert opinion that the process of covering hazards by doing, first, their apportionment/grouping, which is then followed by designing protection system for each group, is a reliable pattern that ensures hazard coverage (claim HC) if the two subprocesses are properly executed. Similarly, if we do not want to consider the final inference SS\implBy HI \& HP as being merely a definition of hazards, we need to reformulate it as  a claim about SEP decomposition: if the processes FHA and HP (system protection against all identified hazards) were properly done, then the system is safe. Clearly, semantic validity of such a claim needs a special justification, which is typically missing from safety cases employing this pattern. 


\subsection{\wfp\ vs. \GSN.}
\label{sec:safe-systems}
We do not think that the goal of a GSN inference tree is to demonstrate  trivial tautologies, and there should be a real content behind the inference steps.
As our discussion above shows, this real content consists of claims about process decomposition -- each inference step is a decomposition step, so that the entire inference tree amounts to justifying the SEP definition. In other words, the actual goal of the blue part of \figinf\ is to ensure the SEP definition given by \figmetamod. To achieve this goal, each decomposition/inference step should be provided by a justification assuring the semantic validity of the step, and we come to item a) in our general definition of safety \defref{main}, which so far was largely ignored in our discussion. Importantly, the \wfp\ formalism per se, including its formal constraint derivation, does not address the issue we are discussing, and the term \wfp\ in the title of this subsection refers to the entire approach to assurance we describe in this report rather than to a special type of workflow modelling. 


A major question we are discussing is this: 
{\em Why \wfsepdef, if properly executed for a system $X$, results in a system $X$ with required critical properties}, 
and the \wfp-approach offers the following generic answer.  
{A proper execution of $\wfsepexe(X)$ results in a system $X$ with the required critical properties because: a) the SEP definition is based on a collective experience in dealing with systems similar to $X$; b) all novelties brought by $X$ were analyzed by experts and addressed according to general patterns of addressing such novelties, and the SEP definition was appropriately amended; c) other arguments justifying that the SEP satisfies its required critical properties. Our formulations have a large room for improvement and accuracy, we give them just to show the nature of the required arguments.  
} 


In the \wfp, assurance conditions such as a,b,c above, are subsumed by an appropriately defined notion of conformance of the SEP definition workflow \wfsepdef\ to some virtual workflow \wfnorm, which models and integrates the set of all relevant  normative and guiding (formal and informal) documents into a workflow \wfnorm\ (the next section provides some detail). 
Thus, Parts 1 and  1* of a generic safety case can be combined into 

{\bf Part 1$^+$:} Assuring that the SEP is properly defined by showing its conformance (as a workflow \wfsepdef) to some normative workflow \wfnorm\ (both are actually \wfp-objects).

In general, this conformance can be managed similarly to the instance conformance we considered in this report (Sect. 3-4). The next section provides some details. 
\renewcommand\papertr[2]{#2}
\subsection{Conformance to Normative doc}\label{sec:conformToNorm}
\papertr{
\begin{figure}
\centering
    \includegraphics[
                    width=0.95\columnwidth%
                 ]%
                 {figures/normative}
\caption{A sample  normative document fragment  \label{fig:normative}}
\vspace{-.4cm}
\end{figure}

}{
\begin{figure*}
\centering
    \includegraphics[
                    width=0.95\textwidth%
                 ]%
                 {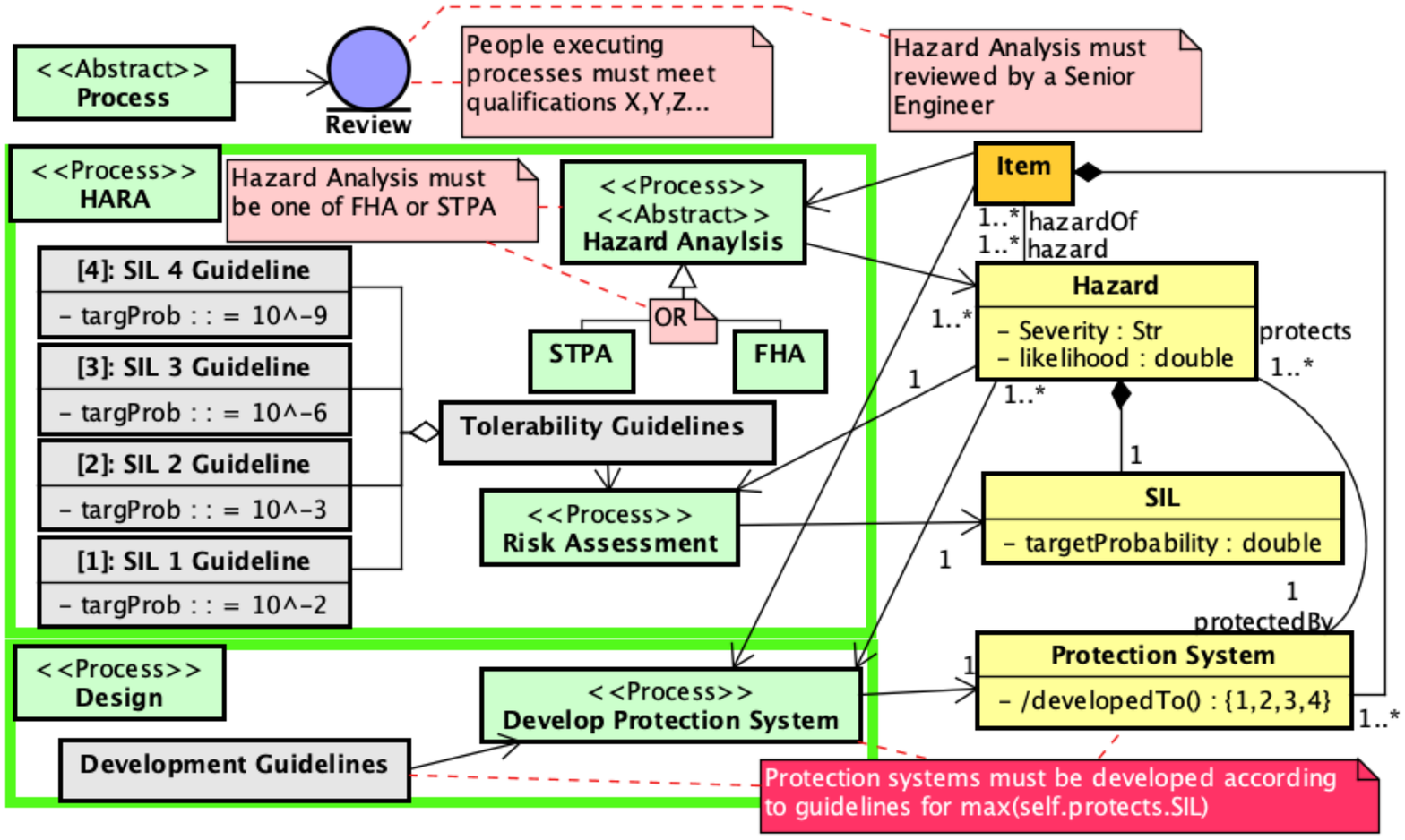}
\caption{A sample  normative document fragment  \label{fig:normative}}
\end{figure*}

}
\subsubsection{The Normative Document}
The main question above is also discussed in paper \cite{alan-caseAgainst-10}, which outlines how the issue is addressed in the classical engineering domains. \paperGM{Similar to our \req\ \textbf{R6}, one}{One} of their main points  is that  sufficiently prescriptive guidelines based on established foundations (we say `normative documents')  for the devlopment of safety-critical systems must be provided to ensure their safety.
However, the question of whether the SEP definition conforms to the \corring\ established foundations/normative docs is far non-trivial. Traditionally, the latter are written in natural languages, contain many cross referenced sections, subsections, tables, \etc, and their full understanding can require a huge amount of time and efforts. Moreover, even full understanding of a standard (such as ISO 26262) does not guarantee its uniqueness because of ambiguities inherent in natural language formulations and possibly different backgrounds of the standard writers and the standard readers. 

In \wfc, we propose to model the integrity of normative documents (domain-specific standards, internal company procedures, legal requirements) as a metamodel \wfnorm\ to guide the development of a \wfsep\ and, moreover, to provide a definitive second layer of assurance by checking conformance $\wfsep\, \confcheck{\reff}{} \,\wfnorm$ (see item a) in \defref{mainF}).  A manufacturer would develop their \wfsep\ by refining a very abstract \wfnorm\ to the point that it can be executed. Through conformance to \wfnorm, a manufacturer can be sure that they satisfy all prescriptive guidelines and thus gain confidence in the safety of their system (as long as \wfnorm\ is reliable indeed). 

Similarly to the instance-metamodel conformance, the \refin\ conformance also has two components: syntactic and semantic. Syntactic conformance is shown by demonstrating that a SEP metamodel is indeed a refinement of the normative metamodel in some formal sense. Semantic conformance should validate that 
the SEP satisfies the intent of the normative metamodel, and is much more challenging. We propose to address it similarly to the issue of SEP execution validity, \ie, by reviewing: each piece of the SEP definition should be checked for conformance to the normative docs and then reviewed. 

\nanewok{From an MDE perspective, replacing the traditional approach with a model-based \wfp\ could provide the following benefits:  a) it avoids the ambiguity of natural language; b) it improves understandability by providing explicit process flow, dataflow, and constraints; c) it presents information more intuitively by utilizing the workflow notations.}{}

\papertr{
	 
}{
\begin{remark}[Multiple Normative Documents]
Manufacturers normally must conform to many domain-specific standards, internal company procedures, legal requirements, and other normative documents. These standards are often interrelated, overlap and can even be contradictory. The \wfp-approach models this as a single virtual  workflow \wfnorm\  defined as an imaginary merge of the components, in which discrepancies are reconciled. Implementation of such a merge requires an accurate specification of all complex relationships between normative documents, and the \corring\ tool support for model matching and merging. 
\end{remark}
\begin{remark}[Deeper justification]\label{sec:reasandjust}
As explained in \sectref{gsn2wf2}, establishing the soundness of an argument may be difficult.  While an instance must only demonstrate conformance to its SEP, and an SEP must only demonstrate conformance to its normative documents/workflow, the latter must also be justified somehow. This justification can be interpreted as yet another one conformance mapping to a system of informal elements based on experience, empirical evidence, best practices, and patterns. 
\end{remark}
}


\section{Towards a UML profile for \wfp\ modelling}
\label{sec:umlprofile}

\begin{figure}
\centering
    \includegraphics[
                    width=0.5\textwidth%
                 ]%
                 {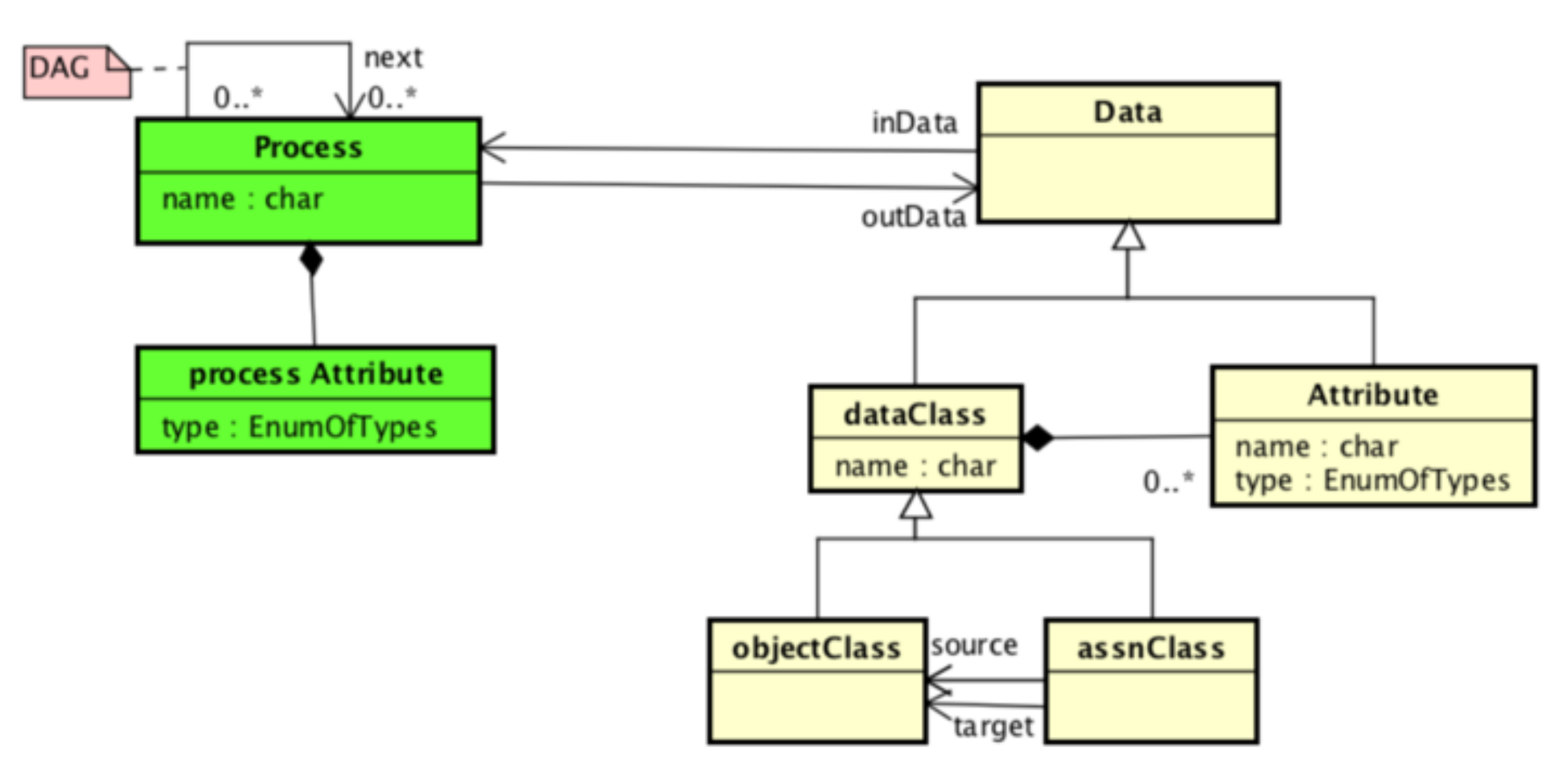}
\caption{Metamodel for \wfp  \label{fig:wfp-metamodel}}
\end{figure}

\newcommand\wfmmfig{\figref{fig:wfp-metamodel}}

\subsubsection{The meta-metamodel.}The \wfp-models we used in the paper are encoded as UML class diagrams of a special type. Special features of this type are
\begin{enumerate}
	\item Two types of classes: process classes and data classes
	\item Two types of associations:
	\begin{enumerate}
		\item dataflow associations from data to process classes and back
		\item static data associations from data to data classes. Specifically, there are associations from the output data of a process to its input data -- this is the essence of \wfp.   
	\end{enumerate}
\item Special process class: Reviewing. Each data elements is supplied with mandatory validity checking. The latter is realized as a cross-cutting concern (a.k.a as an aspect), \ie, a metamodel fragment attached to the main metamodel at a specified point as explain in Sect.3.4 
  \item Several constraints regulating the interaction above. The most important one is that the process classes and their dataflow form an hierarchy (a directed   acyclic graph without looping, DAG); note that looping mentioned in 1b is purely static!
\end{enumerate}

Technically, the requirements above can be implemented as a UML profile, or a domain specific language. A high-level meta-metamodel for this profile is shown in \wfmmfig. 


\subsubsection{\wfp\ Design.} \wfp-design methodology,as any design methodology, is difficult to formalize, but we have found several patterns demonstrated in Sect.3. We start with a GSN case and convert it into a \wfp\ by following the following steps.
\begin{enumerate}
\item Label the GSN diagram with processes involved (green labels in Fig.4) The SEP-logic may require adding new (sub)processes.
\item Discover the dataflow of the processes identified in 1 (the yellow and grey labels in Fig.4). 
\item Identify claims and model them as constraints with Boolean truth-values (red labels). 
\item Supply each data associations with a reviewing block (violet circles in Fig.6) thus converting fundamental validity requirements of a principally semantic nature into syntactical constraints. 
\item Add logical claim inference -- blue blocks and blue arrows in Fig.7. 
\end{enumerate}

\subsubsection{Model management.} There are two special requirements to a model management engine to handle \wfp-models. 
\begin{enumerate}
\item To execute aspect weaving, \ie, be able to add the reviewing metamodel to the main metamodel at any specified point.

\item To have a special query interface for the special structure of the metamodel, \eg, build the control flow view of a \wfp-model (by transitive closure of the In-Out relationships), or build a data dependency graph w.r.t. the dataflow embodied into the \wfp-model.
\end{enumerate}

{  
\paperGM{
	
\section{Applicability and Possible Impact of \wfp
}\label{sec:compara-survey} 

\newcommand\gsnbf{{\bf\em GSN}}
\newcommand\citsur[1]{\em\small #1}

We have analyzed \paperGM{\papertr{five}{seven}}{\papertr{five}{seven}} major issues with SCs identified in a survey on SCs used in practice \cite{jcheng-corr18}. These issues correspond well with our own experience of SCs in industry\paperGM{and overlap with our requirements described on p.1}{}. Below, for each issue we provide its italicized summary
\paperGM{(and refer to our similar requirements),}{}
\papertr{followed by how \wcfa\ corroborates and/or addresses the issue.
In the accompanying TR, we also discuss GSN's inability to manage (or, worse, \awmodok{}{its}{} ability to create) the issues identified in the survey. 
}{
followed by the applicability of \wcfa\ in corroborating and/or addressing the issue, followed by GSN's \awmodok{inapplicability}{inability}{} to manage (or, worse, \awmodok{}{its}{} ability to create) the issue.
}

\renewcommand\citsur[1]{{\small\em #1}}
\renewcommand\subsubsection[1]{\item #1}
 \begin{enumerate}[label={\bf  \arabic*)}, 
	wide=0pt]
\subsubsection{\textbf{Scalability}}
\citsur{Navigation and comprehension of the SC is difficult. Current tools either fail to deliver features to handle scalability, or are domain-specific and not generalizable. \paperGM{(Plus R4, R5)}{}}   


\papertr{
All traceability is explicit and extends into the development process, the product artifacts and even the environment, thereby improving detailed understandability. The \fwk\ is model-based and  allows for compositionality and hence decomposition and modularization and so, in spite of the complexity of the system and its trace-links, there are built in constructs for managing this.
}
{\nanewok{\wfbf: a) All traceability is explicit and extends into the development process, the product artifacts and even the environment, thereby improving detailed understandability, b) The \fwk\ is model-based and  allows for compositionality and hence decomposition and modularization and so, in spite of the complexity of the system and its trace-links, there are built in constructs for managing this.}{}
	
\gsnbf: SCs are difficult to comprehend due to the amount of implicit information and their ad hoc structure that is evident even in simple SCs (see \figref{fig:GSNannotated-full}); for large industrial SCs, these deficiencies can accumulate enormously and really prevent understanding. 
}

\subsubsection{\textbf{Managing Change}}
\citsur{There are no effective mechanisms to manage changes in SCs. Lack of integration of processes and inadequate tool support are contributing factors. 
	\paperGM{(R1, R2, R5)}{}
} 

\papertr{
\awmodok{Complete}{The SEP and SC meta-models are integrated. We thus have complete}{} traceability between model elements, explicit dataflow and explicit process flow, all of which are crucial for impact analysis and change management. A strict separation of instance and \awmodok{meta- (and deeper if needed) levels}{meta-levels}{}  of modelling provides a convenient reusability mechanism. 
}
{
\wfbf: \awmodok{Complete}{The SEP and SC meta-models are integrated. We thus have complete}{} traceability between model elements, explicit dataflow and explicit process flow, all of which are crucial for impact analysis and change management. A strict separation of instance and \awmodok{meta- (and deeper if needed) levels}{meta-levels}{}  of modelling provides a convenient reusability mechanism. 

\gsnbf: Traceability is almost entirely omitted,
making change management a very difficult task. \awmodok{Moreover, the lack of well defined semantics of syntactic elements, 
makes the development of tools supporting automatic change management very difficult.}{The GSN SC appears as entity separate from the SEP and other product artifacts.}{}
}

\subsubsection{\textbf{Requiring special skills to create}}
\citsur{Graphic notations of SCs are  intuitive to understand, but creating a well-structured, convincing SC requires special skills/experience. The privacy of information in SCs prevents the sharing of knowledge on effective patterns and strategies. \paperGM{(R3: a unified approach reduces the need for special skills.)}{}
} 

\papertr{
Workflows are natural for engineers and \wcfa\ may  provide an intuitive and easily understandable framework for building SCs. A strict separation of instance and (meta-)metalevels  of modelling allows experts to create templates and share effective patterns and strategies \zdmoddno{(with a reduced fear of revealing sensitive information about products).}{(and the fear of revealing sensitive information about products is essentially mitigated.}{}{-3em}
}
{
\wfbf: a) Workflows are a natural way for engineers to understand their work. As a result, \wcfa\ provides an intuitive and easily understandable framework for building SCs. b) A strict separation of instance and meta- (and metameta-) levels  of modelling allows experts to create templates and share effective patterns and strategies \zdmoddno{(with a reduced fear of revealing sensitive information about products).}{(and the fear of revealing sensitive information about products is essentially mitigated.}{}{-3em}

\gsnbf: For engineers, it is difficult to convert their intuitive understanding of the story of a system's safety into the one-dimensional structure of GSN-style safety cases.
}

\subsubsection{\textbf{Complexity of the system}}
\citsur{Identifying safety concerns is extremely difficult in complex systems due to the difficulty of managing and identifying interactions between different subsystems. Multidisciplinary collaboration is also challenging. 
	\paperGM{(R2, R4, R5)}{}
} 

\papertr{
\awmodok{On complexity.}{Complexity.}{} \zdmoddok{WF+ remedies this issue by a)  including the traceability required to manage complex interactions and b) supporting the modularity necessary for humans to be able to interpret complex structures.}
 Traceability does help to identify interactions. Compositionality of \wfp\ allows for decomposition a classical engineering approach to manage complexity.

\awmodok{On collaboration.}{Collaboration.}{} An intuitive common language allows experts to share different perspectives and facilitate discussions
}
{
\awmodok{On complexity.}{Complexity.}{} \wfbf: \zdmoddok{WF+ remedies this issue by a) including the traceability required to manage complex interactions and b) supporting the modularity necessary for humans to be able to interpret complex structures.}
{a) Traceability does help to identify interactions. b) Compositionality of \wfp\ allows for decomposition a classical engineering approach to manage complexity.}
{just to show how it works}{-5em}

\gsnbf:\zdmoddok{
GSN-style SCs do not support the traceability necessary to effectively analyze interactions between different components unless the reader was deeply involved in the creation of this system. 
}{The absence of traceability links makes interaction analysis very difficult unless the SC assessor has been deeply involved in the creation of this system. But then the assessment can suffer from confirmation bias. 
}{}{-5em}

\awmodok{On collaboration.}{Collaboration.}{} \wfbf: An intuitive common language allows experts to share different perspectives and facilitate discussions

\gsnbf: Ad-hoc structure and implicit information leaves room for ambiguity, potentially impeding communication between teams.
}


\subsubsection{\textbf{Uncertainty, trust, confidence}}
\citsur{\small Issues of confidence and trust are seen as intangible. This leaves room for doubt about the efficacy of SCs and makes it difficult to establish trust. \paperGM{(R3, R4, R6)}{}
} 


\nanewok{Confidence and trust are notoriously difficult to quantify in the context of assurance, and have long been a topic of debate \citedebate.}{}
\papertr{
 Conformance of the SEP-definition to normative documents makes a  ``half'' of the \fwk\ and thus directly addresses the issues above. 
 Creating data-driven argument flow in \wfp\ is difficult, 
but if managed, it provides a high level of confidence as it is explicit and formalized.
}
{
\wfbf:  a) Conformance of the SEP-definition to normative documents makes a  ``half'' of the \fwk\ and thus directly addresses the issues above. 
b) Creating data-driven argument flow in \wfp\ is difficult, 
but if managed, it provides a high level of confidence as it is explicit and formalized.

\gsnbf: The ability of manufacturers to generate trust in their safety cases is dependent upon their ability to express what has been done to reviewers. GSN-style SCs provide little support for this as they leave massive arrays of data, process and dataflow details implicit. 
}
\papertr{
\end{enumerate}
 
}
{
\subsubsection{\textbf{Too flexible}}
\citsur{\small Flexibility can be dangerous because manufacturers can overlook aspects of safety. Conflicts of interest and confirmation bias contribute to this danger. This is closely related to establishing trust and confidence. 
\paperGM{(R6)}{}
}

Independent assessment has been shown to be effective in reducing confirmation bias \cite{levesonSCinCertif}.
\papertr{
A well-defined structure for SCs and the ability to offer templates to guide their development, allowing assessors to gain expertise in evaluating SCs. 
}
{
\wfbf: A well-defined structure for SCs and the ability to offer templates to guide their development, allowing assessors to gain expertise in evaluating SCs. 

\gsnbf: The ad-hoc structure of GSN-style safety cases makes it very difficult to develop expertise in analyzing these safety cases, making it far too likely that holes in arguments will go unnoticed. 
}

\subsubsection{\textbf{Incomplete information}}
\citsur{Survey: Limited integration of SC management into the software development process causes challenges in collecting sufficient and accurate information for safety arguments. Incomplete traceability, insufficiencies in testing and flawed safety requirements are all issues that arise as a result of this. 
\paperGM{(R3, R4, R5)}{}
} 

\papertr{
Processes and complete traceability between data are incorporated. This enables integration of the Design and SEP and SC development, allowing for systematic change impact analysis and, with proper tooling, automated updating of SCs in response to changes in data and gathering of data, ensuring information is properly managed.
}
{
\wfbf: Processes and complete traceability between data are incorporated. This enables integration of the Design and SEP and SC development, allowing for systematic change impact analysis and, with proper tooling, automated updating of SCs in response to changes in data and gathering of data, ensuring information is properly managed.

\gsnbf: a) GSN-style safety cases argue over instance-level data. The notion of processes flow and dataflow are omitted almost entirely. b) Implicit traceability leaves room for data to be left behind or forgotten. 
}
} 
\end{enumerate}

}{}
} 
\paperGM{
\section{Related 
	work}\label{sec:relwork}

\begin{wrapfigure}{R}{0.45\columnwidth}
\centering
    \includegraphics[
                    width=0.45\columnwidth%
                 ]%
                 {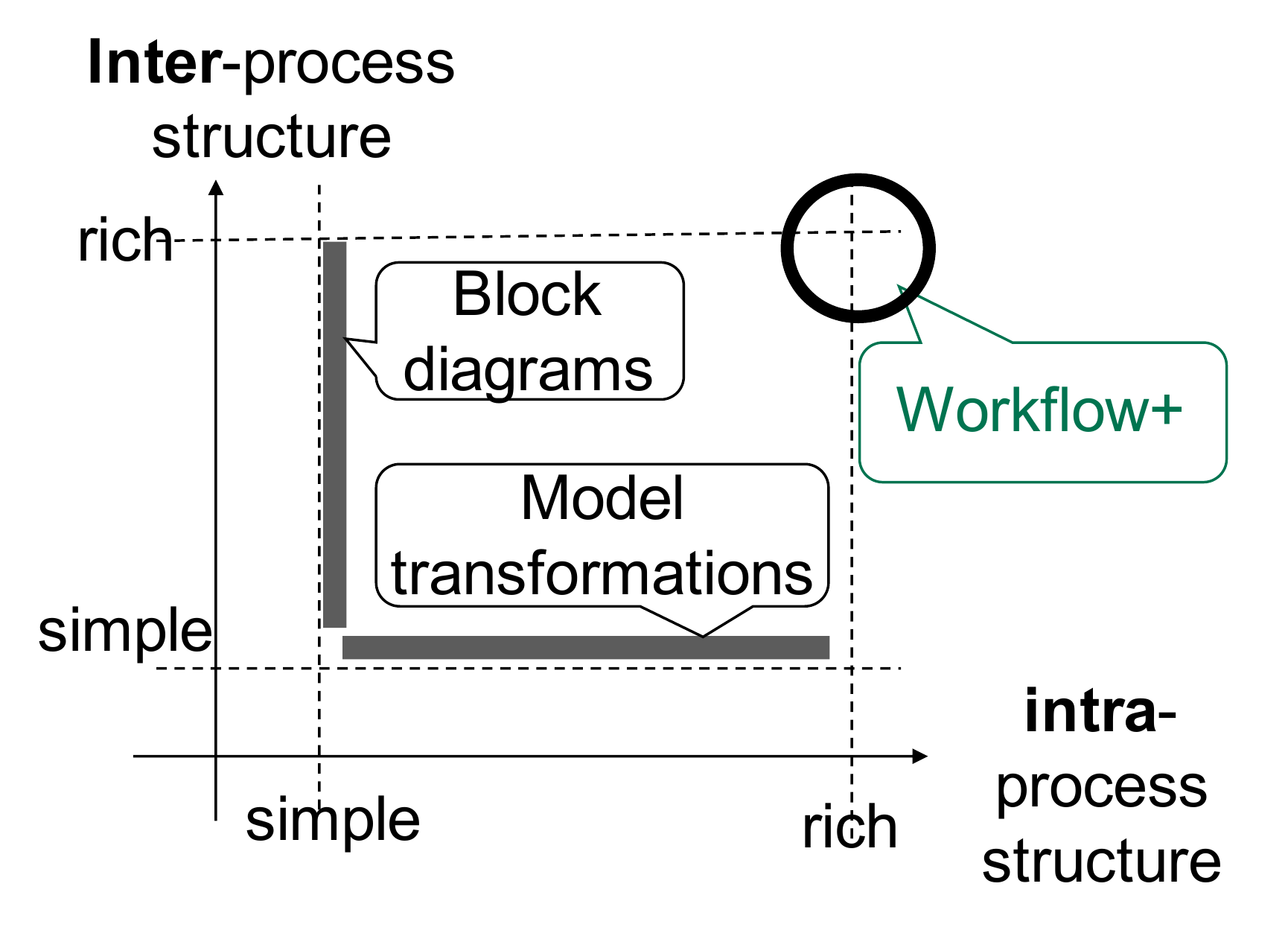}
                 \vspace{-7mm}
\caption{}
\label{fig:wfPlane}
\end{wrapfigure}
A series of seminal papers \cite{vdbrand-icsr13,lionel-ist14,sahar-models16} introduced the ideas of metamodeling and model management into assurance, but their main focus was on data conformance rather than dataflow. \awmodok{The latter is in focus of paper}{Dataflow was the focus in}{} \cite{me-models18} but they do not consider the argument \awmodok{flow there,}{flow,}{} and focus on process decomposition rather than on SCs. Analysis of GSN-based SCs has an extensive literature, but it is mainly focused on the 
logic than the dataflow (see a survey in  
\cite{RushbySRI2015}). Several ways of formalizing SCs were suggested by Rushby in \cite{rushby2010formalism}, but his focus is more on the argument structure. 
Tooling for GSN-based SCs is discussed in \cite{denney2018}

\wfp\ has its origin in block diagrams (as a networking mechanism) and model \trafon s as processes over GBOO data, but differs from both of them as illustrated by the schema in \figref{fig:wfPlane}. 
Model refinement \cite{refinTK}  and process refinement \cite{?} \awmodok{[look at Thomas Kuhne paper]}{}{} have extensive literature, but we have \awmodok{still not found}{not found yet}{} the notion most suitable for our needs to model conformance to the normative documents.

}{}
\section{Conclusion
}\label{sec:concl}

We have shown that GSN-style safety cases suffer from two inherent drawbacks. The first, technical, is implicitness and incompleteness of the dataflow and traceability in the GSN-structured safety cases (see Sect. 5). The second, conceptual, is the absence of assuring the SEP definition, which is an important constituent of safety assurance (Sect. 6). The worst feature of GSN safety cases is that they fake logical inference and thus create a false sense of confidence that safety had been demonstrated acceptably well. 

We have presented \wfp\ as a modelling framework \awmodok{which, on the specification level,}{which}{} meets the core needs of creating, documenting and managing safety cases. Precise dataflow modelling, the benefits of the traceability inherent in the UML class modelling, the intuitive understandability of workflows, and a well-founded approach to developing assurance claims modelled by OCL constraints over a SEP metamodel, result in an approach that integrates processes, product and safety cases. This integration is both conceptual and technical---the latter is based on explicit traceability links relating all constituents that matter in the SEP. This comprehensive traceability is crucial in consistency management, in which we strive to maintain consistency between the various artifacts within all the associated models.

\bibliographystyle{ieeetran}
\bibliography{fromFase19-2} 
\end{document}